\documentclass[a4paper,10pt,twoside]{cpc-hepnp}
\usepackage{CJK,upgreek,fancyhdr}
\usepackage{multicol}
\usepackage{graphicx}
\usepackage{booktabs}
\usepackage{amssymb,bm,mathrsfs,bbm,amscd}
\usepackage[tbtags]{amsmath}
\usepackage{lastpage}
\usepackage{hyperref}
\hypersetup{colorlinks=true, citecolor=blue, linkcolor=blue,filecolor=black,urlcolor=blue}
\usepackage{color}

\begin{document}
\begin{CJK*}{GB}{gbsn}

%\fancyhead[c]{\small Chinese Physics C~~~Vol. xx, No. x (201x) xxxxxx}
\fancyfoot[C]{\small \thepage}

%\footnotetext[0]{Received 1 September 2017}

    \title{Statistical method in quark combination model \thanks{Supported by National Natural Science Foundation of China (11535012, 11890713, 11575100), the 973 program (2015CB856902), the Key Research Program of the Chinese
Academy of Sciences (XDPB09), and Shandong Province Natural Science Foundation (ZR2019YQ06,ZR2019MA053).  
 We dedicate this work to Qu-bing Xie (1935-2013) who was the teacher,
    mentor and friend of ZTL, FLS and QW.  }}

\author{%
Yang-guang Yang(ÑîÑô¹â)$^{1}$ 
\quad Jun Song (Ëξü)$^{2;1)}$\email{songjun2011@jnxy.edu.cn}
\quad Feng-lan Shao(ÉÛ·ïÀ¼)$^{3;2)}$\email{shaofl@mail.sdu.edu.cn} \\
\quad Zuo-tang Liang(Áº×÷ÌÃ)$^{4;3)}$\email{liang@sdu.edu.cn} 
\quad Qun Wang(ÍõȺ)$^{1;4)}$\email{qunwang@ustc.edu.cn}
}
\maketitle

\address{%
$^1$  Department of Modern Physics, University of Science and Technology of China, Hefei, Anhui 230026, China\\
$^2$  Department of Physics, Jining University, Shandong 273155, China\\
$^3$  School of Physics and Engineering, Qufu Normal University, Shandong 273165, China \\
$^4$  Institute of Frontier and Interdisciplinary Science, \\ Key Laboratory of Particle Physics and Particle Irradiation (MOE), Shandong University, Qingdao, Shandong 266237, China
}

\begin{abstract}
    We present a new method of solving the probability distribution for baryons, antibaryons and mesons at the hadronization of constituent quark and antiquark system. The hadronization is governed by the quark combination rule in the quark combination model developed by the Shandong Group. We use the method of the generating function  to derive the outcome of the quark combination rule, which is much simpler and easier to be generalized than the original method. Furthermore, we use the formula of the quark combination rule and its generalization to study the property of multiplicity distribution of net-protons.  Taking a naive case of quark number fluctuations and correlations at hadronization, we calculate ratios of multiplicity cumulants of final-state net-protons and discuss the potential applicability of quark combination model in studying hadronic multiplicity fluctuations and the underlying phase transition property in relativistic heavy-ion collisions. 

\end{abstract}



%\footnotetext[0]{\hspace*{-3mm}\raisebox{0.3ex}{$\scriptstyle\copyright$}2013
%Chinese Physical Society and the Institute of High Energy Physics
%of the Chinese Academy of Sciences and the Institute
%of Modern Physics of the Chinese Academy of Sciences and IOP Publishing Ltd}%

\begin{multicols}{2}

\section{Introduction\label{sec1}}
Most hadronization models such as the Lund string model \cite{Andersson:1978vj,Andersson:1979ij,Andersson:1983ia,Andersson:1983jt}
or the coalescence/recombination models \cite{Anisovich:1972pq,Bjorken:1973mh,Das:1977cp,Hwa:1979pn,Xie:1988wi,Wang:1994tb,Wang:1995ch,Liang:2000gz,Fries:2003vb,Greco:2003xt,Hwa:2003ic,Hwa:2004ng,Chen:2006vc,Ravagli:2007xx,Miao:2007cm,Ayala:2007cp,Abir:2009sh}
in electron-positron and hadron-hadron collisions assume that a hadron
is formed by quarks and antiquarks in the neighborhood of phase space.
Normally the phase space can be decomposed into the longitudinal and
transverse direction. The longitudinal phase space plays a special
role. For example in the Lund model a string is formed between a quark
and an antiquark moving back to back, it is an object with only one
spatial dimension, i.e. with only the longitudinal phase space. A
quark-antiquark pair is produced as the result of the string break
into two pieces. The process of string breaks continues until it ends
at a lowest energy scale. The transverse momentum space of the excited quark-antiquark pair is rather limited and can be described by the
exponentially suppressed function of the transverse momentum. In relativistic
heavy-ion collisions, the longitudinal phase space is still dominant
although the transverse expansion is significant.

The quark combination model developed by the Shandong group (SDQCM)
\cite{Xie:1985ng,Xie:1988wi,Liang:1991ya,Wang:1994tb,Wang:1995ch,Si:1997ux,Xie:1997ap,Shao:2004cn,Shao:2006sw,Shao:2009uk,Zhao:2010zzg,Song:2010bi,Sun:2011kj,Wang:2012cw,Wang:2013duu,Shao:2015rra,Song:2016ihg}
is a kind of the exclusive or statistical hadronization model which
is different from the Lund string model or the coalescence model.
The model firstly takes the constituent quark degrees of freedom as an effective description for the strongly-interacted quark gluon system at hadronization. Then, the model
 adopts a quark combination rule (QCR) to combine the quarks
and antiquarks in the neighborhood of the longitudinal phase space
into baryons and mesons. Since the longitudinal phase space is easily
described by the momentum rapidity, the correlation in rapidity is
the basis of the QCR. Such
kind of QCR in SDQCM has successfully explained many data of hadronic
production in $e^{+}e^{-}$ and $pp$ collisions \cite{Chen:1988qi,Fang:1989ex,Si:1997ux,Xie:1997ap}
and also rapidity distributions of hadrons in heavy-ion collisions
\cite{Shao:2006sw,Song:2008iq,Shao:2009uk,Sun:2011kj}. 

The probability distributions of the particle numbers of baryons,
antibaryons and mesons is an important observable in high energy collisions
and are closely related to the hadronization dynamics. Especially
the particle number distributions are essential to search for the
critical end point of the QCD phase diagram in relativistic heavy-ion
collisions \cite{Stephanov:1999zu,Asakawa:2000wh,Chen:2014ufa,Aggarwal:2010wy,Adamczyk:2013dal}.
In this paper, we propose a new method to solve the baryon and meson
number distribution in the context of the QCR in the SDQCM. The new
method is much simpler and easier to be generalized to more sophisticated
cases than the original one \cite{Mo:1986cl}. Then we use these
formula to calculate the multiplicity distribution and the ratios
of cumulants for net protons in relativistic heavy-ion collisions. 

The paper is organized as follows. In Sec. \ref{sec:old_QCR}, we
briefly present the original QCR in the SDQCM. Then we use the generating
function method to solve the particle number probability of baryons,
antibaryons and mesons for given numbers of quarks and antiquarks.
In Sec. \ref{sec:new_QCR} we generalize the original QCR and derive
the corresponding particle number probability of baryons, antibaryons
and mesons using the generating function method. In Sec. \ref{sec:compara_QCR}
we compare the numerical difference between the original QCR and the
generalized QCR in terms of moments of the antibaryon number. In Sec.
\ref{sec:id_h_fc}, we formulate the fluctuation and correlation of
identified hadrons. In Sec. \ref{sec:incl_qnfc} and \ref{sec:decay_formulas},
we study the effects of the quark number fluctuation and resonance
decays. In Sec. \ref{sec:Results-and-discussions}, we give an illustrative
example of applying the QCR and generalized QCR to calculate the ratios
of cumulants for net protons in heavy-ion collisions and compare with
data. Finally we give a summary and make discussions in Sec. \ref{sec:summary}.

\section{Baryon and meson formation in original quark combination rule\label{sec:old_QCR}}
Due to the non-perturbative QCD feature, the transition from quarks and/or gluons to hadrons is not yet solved from first principle and is only described by the phenomenological models at present. Inspired by the simple and effective description of constituent quark model in explaining the static property of hadrons, SDQCM assumes the constituent quarks and antiquarks as the effective degrees of freedom for the strongly-interacted quarks and gluons at hadronization and therefore builds a simple hadronization phenomenology by the combination of these constituent quarks and antiquarks into hadrons. We emphasize that there exist explicit experimental signals for such constituent quark degrees of freedom in high energy collisions, such as the quark number scaling property of elliptic flows and transverse momentum spectra for hadrons observed in recent years \cite{Adare:2006ti, Adamczyk:2015ukd, Sirunyan:2018toe, Song:2017gcz, Zhang:2018vyr, Song:2019sez}. A phenomenological quark combination rule was proposed in Ref.~\cite{Xie:1985ng} to describe the combination of these constituent quarks and antiquarks in one-dimensional phase space and can successfully describe the data of yields and momentum distributions of hadrons in high energy reactions \cite{Chen:1988qi,Fang:1989ex,Shao:2006sw,Song:2008iq,Shao:2009uk,Sun:2011kj,Si:1997ux,Xie:1997ap}. In this section, we use the generating function method to solve the probability distribution for the number of baryons, antibaryons and mesons formed by QCR.

\subsection{Original quark combination rule\label{sec:qcr-1}}

In the QCM $N_{q}$ quarks and $N_{\bar{q}}$ antiquarks produced
in an event are put into a queue and then allow them to combine into
hadrons one by one following a rule called the quark combination rule
(QCR)\cite{Xie:1985ng}. QCR is based on the basic property of QCD. 
\textcolor{black}{
    A $q\overline{q}$ pair pair may be in a color octet with a repulsive interaction or a singlet with a attractive interaction.  If $q\overline{q}$ is adjacent in phase space, $q\overline{q}$ will have enough time or opportunity to be in a color state and hadronizes into a meson.
 For a $qq$ pair, it may be in a sextet or an antitriplet. If its nearest neighbor is a $q$ in
phase space, they can hadronizes into a baryon. If the neighbor of $qq$ is a $\overline{q}$, $q\overline{q}$ will win the competition to form a meson and leave a $q$ alone to combine with other quarks and/or antiquarks. This is because
the attraction strength of the singlet for $q\bar{q}$ is two times that of
the antitriplet for $qq$ (via counting color factor in one-gluon exchange case).
}
 The original QCR proposed in Ref. \cite{Xie:1985ng} reads: 
\begin{enumerate}
\item Check if there are partons in the queue. If there are no partons,
the process ends. Otherwise, start from the first parton ($q$ or
$\overline{q}$) in the queue and go to the next step. 
\item Look at the second parton. If there is no second parton, the process
ends. Otherwise, if the baryon number of the second parton in the
queue is different from the first one, i.e. the first two partons
are either $q\overline{q}$ or $\overline{q}q$, they combine into
a meson and are removed from the queue, go to step 1; Otherwise they
are either $qq$ or $\overline{q}\overline{q}$, go to the next step. 
\item Look at the third parton. If there is no third parton, the process
ends. Otherwise, if the third parton is different in baryon number
from the first one, the first and third parton form a meson and are
removed from the queue, go to step 1; Otherwise the first three partons
combine into a baryon or an antibaryon and are removed from the queue,
go to step 1. 
\end{enumerate}
The following example show how the above QCR works
\begin{eqnarray}
 &  & q_{1}\overline{q}_{2}q_{3}q_{4}q_{5}\overline{q}_{6}q_{7}q_{8}\overline{q}_{9}\overline{q}_{10}q_{11}\overline{q}_{12}\overline{q}_{13}\overline{q}_{14}\overline{q}_{15}q_{16}q_{17}q_{18}\overline{q}_{19}\overline{q}_{20}\nonumber \\
 & \rightarrow & M(q_{1}\overline{q}_{2})B(q_{3}q_{4}q_{5})M(\overline{q}_{6}q_{7})M(q_{8}\overline{q}_{9})M(\overline{q}_{10}q_{11})\nonumber \\
    &  & \overline{B}(\overline{q}_{12}\overline{q}_{13}\overline{q}_{14})M(\overline{q}_{15}q_{16})M(q_{17}\overline{q}_{19})M(q_{18}\overline{q}_{20}). \label{exqcr}
\end{eqnarray}
    In relativistic heavy-ion collisions, the longitudinal rapidity space is predominant and the rapidity density of quarks and antiquarks is quite large. Therefore, it is suitable and straightforward to apply such a QCR in one-dimensional longitudinal rapidity space. 
However, it is quite complicated in three dimensional phase space because one can not easily define a particular hadronization order to apply QCR \cite{Hofmann:1999jx}. 

\subsection{Recursive relation for $F(N_{M},N_{B},N_{\bar{B}},N_{r},N_{\bar{r}})$}

We consider the system consisting of $N_{q}$ quarks and $N_{\bar{q}}$
antiquarks stochastically populated in one-dimensional phase space.
After combination by QCR, there are $N_{B}$ baryons, $N_{\bar{B}}$
antibaryons, and $N_{M}$ mesons formed and $N_{r}$ quarks and $N_{\bar{r}}$
antiquarks left. There are only five different configurations with
$(N_{r},N_{\bar{r}})=(0,0),(0,1),$ $(1,0),(2,0),(0,2)$. The quark
number conservation gives 

\begin{align}
N_{M}+3N_{B}+N_{r} & = N_{q},\nonumber \\
N_{M}+3N_{\bar{B}}+N_{\bar{r}} & =  N_{\bar{q}}.\label{eq:sum-n}
\end{align}
The outcome of implementing the QCR to the queue of $N_{q}$ quarks
and $N_{\bar{q}}$ antiquarks gives a group of numbers $(N_{M},N_{B},N_{\bar{B}},N_{r},N_{\bar{r}})$.
The queue $(N_{M},N_{B},N_{\bar{B}},0,0)$ can be reached by one of
the following ways of adding one $q$ or $\bar{q}$ to the end of
other four queues with smaller number of quarks and antiquarks 
\begin{align}
(a) &   (N_{M}-1,N_{B},N_{\bar{B}},1,0)+\bar{q},\nonumber \\
(b) &   (N_{M}-1,N_{B},N_{\bar{B}},0,1)+q,\nonumber \\
(c) &   (N_{M},N_{B}-1,N_{\bar{B}},2,0)+q,\nonumber \\
(d) &   (N_{M},N_{B},N_{\bar{B}}-1,0,2)+\bar{q}.\label{eq:add-rule}
\end{align}

We use $F(N_{M},N_{B},N_{\bar{B}},N_{r},N_{\bar{r}})$ to denote the
number of different queues for a given group $(N_{M},N_{B},N_{\bar{B}},N_{r},N_{\bar{r}})$.
We define $F(0,0,0,0,0)=1$ and $F(N_{M},N_{B},N_{\bar{B}},N_{r},N_{\bar{r}})=0$
for the case that any of $N_{M}$, $N_{B}$, $N_{\bar{B}}$, $N_{r}$
and $N_{\bar{r}}$ are negative. Under constraint Eq. (\ref{eq:sum-n}),
the sum of $F(N_{M},N_{B},N_{\bar{B}},N_{r},N_{\bar{r}})$ over all
different groups of $(N_{M},$ $N_{B},N_{\bar{B}},N_{r},N_{\bar{r}})$
should be 
\begin{align}
    S(N_{q},N_{\bar{q}}) = & \sum_{\{N_{M},N_{B},N_{\bar{B}},N_{r},N_{\bar{r}}\}}F(N_{M},N_{B},N_{\bar{B}},N_{r},N_{\bar{r}})\nonumber \\
   & \times\delta_{N_{M}+3N_{B}+N_{r},N_{q}}\delta_{N_{M}+3N_{\bar{B}}+N_{\bar{r}},N_{\bar{q}}}\nonumber \\
 & =  \left(\begin{array}{c}
N_{q}+N_{\bar{q}}\\
N_{q}
\end{array}\right)\equiv\frac{(N_{q}+N_{\bar{q}})!}{N_{q}!N_{\bar{q}}!},\label{eq:snq-nqb}
\end{align}
where $\delta_{i,j}=1$ if $i=j$ and $\delta_{i,j}=0$ if $i\neq j$. 

For non-zero $N_{r}$ and $N_{\bar{r}}$, $F(N_{M},N_{B},N_{\bar{B}},N_{r},N_{\bar{r}})$
has a property 

\begin{equation}
F(N_{M},N_{B},N_{\bar{B}},N_{r},N_{\bar{r}})=\sum_{i=0}^{N_{M}}F(i,N_{B},N_{\bar{B}},0,0).\label{eq:lemma1}
\end{equation}
Proof of Property. We firstly take $(N_{r},N_{\bar{r}})=(1,0)$ for
example. We note that the queue giving $(N_{M},N_{B},N_{\bar{B}},1,0)$
can be reached by two ways: (a) add $q$ to the end of the queue with
$(N_{M},N_{B},N_{\bar{B}},0,0)$; (b) add $\bar{q}$ to the end of
the queue with $(N_{M}-1,N_{B},N_{\bar{B}},2,0)$ which can further
be obtained from the queue with $(N_{M}-1,N_{B},N_{\bar{B}},1,0)$
by adding a $q$ to the end. Thus we obtain the recursive relation
\begin{align}
 & F(N_{M},N_{B},N_{\bar{B}},1,0)\nonumber \\
 & =F(N_{M},N_{B},N_{\bar{B}},0,0)+F(N_{M}-1,N_{B},N_{\bar{B}},2,0)\nonumber \\
 & =F(N_{M},N_{B},N_{\bar{B}},0,0)+F(N_{M}-1,N_{B},N_{\bar{B}},1,0).
\end{align}
Solving the above equation recursively we get Eq. (\ref{eq:lemma1}),
where we have used $F(0,N_{B},N_{\bar{B}},1,0)=F(0,N_{B},N_{\bar{B}},0,0)$.
The proof of the case $(N_{r},N_{\bar{r}})=(2,0)$ is straightforward
due to the fact that the queue giving $(N_{M},N_{B},N_{\bar{B}},2,0)$
can be only obtained from the queue with $(N_{M},N_{B},N_{\bar{B}},1,0)$
by adding a $q$ to the end. The proof for the cases $(N_{r},N_{\bar{r}})=(0,1)$
and $(N_{r},N_{\bar{r}})=(0,2)$ is similar. 

Using properties Eq. (\ref{eq:lemma1}) and Eq. (\ref{eq:add-rule}),
we obtain 

\begin{align}
 & F(N_{M},N_{B},N_{\bar{B}},0,0)\nonumber \\
 & =F(N_{M}-1,N_{B},N_{\bar{B}},1,0)+F(N_{M}-1,N_{B},N_{\bar{B}},0,1)\nonumber \\
 & +F(N_{M},N_{B}-1,N_{\bar{B}},2,0)+F(N_{M},N_{B},N_{\bar{B}}-1,0,2).\label{eq:f00-nm}
\end{align}
We make replacement \textbf{$N_{M}\rightarrow N_{M}-1$} in the above
equation and take the difference between the two, and finally we use
Eq. (\ref{eq:lemma1}) to get the recursive equation for $F(N_{M},N_{B},N_{\bar{B}},0,0)$
\begin{align}
 & F(N_{M},N_{B},N_{\bar{B}},0,0)\nonumber \\
 & =3F(N_{M}-1,N_{B},N_{\bar{B}},0,0)\nonumber \\
 & +F(N_{M},N_{B}-1,N_{\bar{B}},0,0)+F(N_{M},N_{B},N_{\bar{B}}-1,0,0).\label{eq:recursive-f00}
\end{align}
where $N_{M}\geqslant0$, $N_{B}\geqslant0$, $N_{\bar{B}}\geqslant0$
excluding two cases $(N_{M},N_{B},N_{\bar{B}},N_{r},N_{\bar{r}})=(0,0,0,0,0),(1,0,0,0,0)$
which we have $F(0,0,0,0,0)=1$ by definition and $F(1,0,0,0,0)=2$
by simple counting. 

\subsection{Solution to recursive equation by generating function method}

First we consider a special simple case: $N_{M}>1$, $N_{B}=N_{\bar{B}}=0$,
$N_{r}=N_{\bar{r}}=0$, Eq. (\ref{eq:recursive-f00}) is simplified
to 
\begin{equation}
F(N_{M},0,0,0,0)=3F(N_{M}-1,0,0,0,0),
\end{equation}
which immediately leads to the solution 
\begin{equation}
F(N_{M},0,0,0,0)=2\times3^{N_{M}-1},\label{eq:fm}
\end{equation}
for $N_{M}\geqslant1$ with $F(1,0,0,0,0)=2$.

Now we consider the general case for $F(N_{M},N_{B},N_{\bar{B}},0,0)$
with $N_{M}\geqslant0$, $N_{B}\geqslant0$, $N_{\bar{B}}\geqslant0$.
We define the generating function 
\begin{align}
 & A\left(x,y,z\right)\nonumber \\
 & =\sum_{N_{M}=0}^{\infty}\sum_{N_{B}=0}^{\infty}\sum_{N_{\bar{B}}=0}^{\infty}F(N_{M},N_{B},N_{\bar{B}},0,0)x^{N_{M}}y^{N_{B}}z^{N_{\bar{B}}},\label{eq:full_gen_func}
\end{align}
and $F(N_{M},N_{B},N_{\bar{B}},0,0)$ is the coefficient of $x^{N_{M}}y^{N_{B}}z^{N_{\bar{B}}}$
in the polynomial expansion of $A(x,y,z)$ once solved. 

To solve $A(x,y,z)$, we define a partial generating function 
\begin{equation}
A(x;N_{B},N_{\bar{B}})=\sum_{N_{M}=0}^{\infty}F(N_{M},N_{B},N_{\bar{B}},0,0)x^{N_{M}}.\label{eq:part_gen_func}
\end{equation}
Inserting Eq. (\ref{eq:recursive-f00}) for $N_{M}\geqslant1$, we
get 
\begin{align}
 & \left(1-3x\right)A(x;N_{B},N_{\bar{B}})-F(0,N_{B},N_{\bar{B}},0,0)\nonumber \\
 & =A(x;N_{B}-1,N_{\bar{B}})+A(x;N_{B},N_{\bar{B}}-1)\nonumber \\
 & -F(0,N_{B}-1,N_{\bar{B}},0,0)-F(0,N_{B},N_{\bar{B}}-1,0,0).\label{eq:gen-func-1}
\end{align}
In a special case with $N_{B}=0$ we obtain 
\begin{align}
A(x;0,N_{\bar{B}}) & =  \frac{1}{(1-3x)^{N_{\bar{B}}}}A(x;0,0)\nonumber \\
 & =  \frac{2x}{(1-3x)^{N_{\bar{B}}+1}}+\frac{1}{(1-3x)^{N_{\bar{B}}}},\label{eq:a-nb0}
\end{align}
where we have used Eq. (\ref{eq:fm}) and assume $3|x|<1$. In the
same way, we obtain 
\begin{equation}
A(x;N_{B},0)  =  \frac{2x}{(1-3x)^{N_{B}+1}}+\frac{1}{(1-3x)^{N_{B}}}.\label{eq:a-nbar0}
\end{equation}
We also obtain two sum rules 
\begin{align}
\sum_{N_{B}=0}^{\infty}A(x;N_{B},0)y^{N_{B}} & =  \frac{1-x}{1-3x-y},\label{eq:a-nbar-sum}\\
\sum_{N_{\bar{B}}=0}^{\infty}A(x;0,N_{\bar{B}})z^{N_{\bar{B}}} & =  \frac{1-x}{1-3x-z},\label{eq:a-nb-sum}
\end{align}
where the convergent region is $|y|,|z|<|1-3x|$. 

We now multiply Eq. (\ref{eq:gen-func-1}) by $y^{N_{B}}z^{N_{\bar{B}}}$
and take a sum over $N_{B}$ from 1 to infinity and $N_{\bar{B}}$
from 1 to infinity. By noticing $A(x,y,z)=\sum_{N_{B}=0}^{\infty}\sum_{N_{\bar{B}}=0}^{\infty}A(x;N_{B},N_{\bar{B}})y^{N_{B}}z^{N_{\bar{B}}}$
and using Eqs. (\ref{eq:a-nbar-sum}) and (\ref{eq:a-nb-sum}), we
solve the generating function as
\begin{equation}
A(x,y,z)  =  \frac{1-x}{1-3x-y-z}.\label{eq:generating-func}
\end{equation}

To obtain $F(N_{M},N_{B},N_{\bar{B}},0,0)$, we firstly extract the
coefficient of $z^{N_{\bar{B}}}$ 
\begin{equation}
C(z^{N_{\bar{B}}})=\frac{1-x}{(1-3x-y)^{N_{\bar{B}}+1}}.\label{eq:coeff-z}
\end{equation}
Then we extract the coefficient of $y^{N_{B}}$ in $C(z^{N_{\bar{B}}})$
as 
\begin{equation}
C(z^{N_{\bar{B}}}y^{N_{B}})  =  \left(\begin{array}{c}
N_{B}+N_{\bar{B}}\\
N_{B}
\end{array}\right)\frac{1-x}{(1-3x)^{N_{B}+N_{\bar{B}}+1}},\label{eq:coeff-yz}
\end{equation}
where we have used 
\begin{equation}
(1\pm w)^{-n}=\sum_{k=0}^{\infty}\left(\begin{array}{c}
n-1+k\\
k
\end{array}\right)(\mp1)^{k}w^{k},
\end{equation}
for $n>0$ and $|w|<1$. Finally we extract the coefficient of $x^{N_{M}}$
in $C(z^{N_{\bar{B}}}y^{N_{B}})$, i.e., 
\begin{align}
 & F(N_{M},N_{B},N_{\bar{B}},0,0)=3^{N_{M}}\left(\begin{array}{c}
N_{B}+N_{\bar{B}}\\
N_{B}
\end{array}\right)\times\nonumber \\
 & \left[\left(\begin{array}{c}
N_{M}+N_{B}+N_{\bar{B}}\\
N_{M}
\end{array}\right)-\frac{1}{3}\left(\begin{array}{c}
N_{M}-1+N_{B}+N_{\bar{B}}\\
N_{M}-1
\end{array}\right)\right],\label{eq:f-final-1}
\end{align}
which give the final result for 
\begin{align}
 & F(N_{M},N_{B},N_{\bar{B}},0,0)\nonumber \\
 & =\begin{cases}
\frac{(2N_{M}+3N_{B}+3N_{\bar{B}})(N_{M}+N_{B}+N_{\bar{B}}-1)!}{N_{M}!N_{B}!N_{\bar{B}}!}3^{N_{M}-1} & \text{for}\ N_{M}>0,\\
\left(\begin{array}{c}
N_{B}+N_{\bar{B}}\\
N_{B}
\end{array}\right) & \text{for}\ N_{M}=0.
\end{cases}\label{eq:final-result}
\end{align}
Eq. (\ref{eq:final-result}) is the result of QCR in Sect. \ref{sec:qcr-1}
and was first given in Ref. \cite{Mo:1986cl} by\textbf{ }mathematical
induction and traditional combination method. The current derivation
is a simplified one based on the recursive equation and the method
of generating functions. The probability of forming $N_{M}$ mesons,
$N_{B}$ baryons and $N_{\bar{B}}$ antibaryons in a quark system
with $N_{q}$ quarks and $N_{\bar{q}}$ antiquarks is given by 
\begin{equation}
P(N_{M},N_{B},N_{\bar{B}})=\frac{F(N_{M},N_{B},N_{\bar{B}},0,0)}{\left(\begin{array}{c}
N_{q}+N_{\bar{q}}\\
N_{q}
\end{array}\right)}.\label{eq:prob-nm-nb-nbbar}
\end{equation}
Here we have removed the label '0,0' from $P(N_{M},N_{B},N_{\bar{B}})$
since this is the real case in hadronization. 

\section{Baryon and meson formation in generalized quark combination rule\label{sec:new_QCR}}

\subsection{Generalized quark combination rule}

The ratio of baryons to mesons (B/M) given by QCR in Sect. \ref{sec:qcr-1}
is larger than observation of heavy-ion collisions \cite{Wang:1996jy,Xie:1987tg}.
In order to suppress the B/M ratio, we can generalize the QCR in Sect.
\ref{sec:qcr-1} to decrease the formation probability of baryons
relative to mesons. The generalized QCR (gQCR) reads: 
\begin{enumerate}
\item Check if there are partons in the queue. If there are no partons,
the process ends. Otherwise, start from the first parton and go to
the next step. 
\item Look at the second parton. If there is no second parton, the process
ends. Otherwise, if the baryon number of the second parton is different
from the first one, i.e. the first two partons are either $q\overline{q}$
or $\overline{q}q$, they combine into a meson and are removed from
the queue, go to step 1; Otherwise they are either $qq$ or $\overline{q}\overline{q}$,
go to the next step. 
\item Look at the third parton. If there is no third parton, the process
ends. Otherwise, if the baryon number of the third parton is different
from the first one, the first and third parton form a meson and are
removed from the queue, go to step 1; Otherwise the first three partons
are either $qqq$ or $\overline{q}\overline{q}\overline{q}$, go to
next step. 
\item Look at the fourth parton. If there is no fourth parton, the first
three partons form a baryon or an antibaryon and the process ends.
Otherwise, if the baryon number of the fourth parton is different
from the first one, the first and fourth parton form a meson and are
removed from the queue, go to step 1; Otherwise the first three partons
combine into a baryon or an antibaryon and are removed from the queue,
go to step 1. 
\end{enumerate}
The outcome of implementing the gQCR to the queue of stochastically
populated $N_{q}$ quarks and $N_{\bar{q}}$ antiquarks gives a group
of numbers $(N_{M},N_{B},N_{\bar{B}},N_{r},N_{\bar{r}})$. Same as
in the QCR case, the special queue with $(N_{M},N_{B},N_{\bar{B}},0,0)$
can be reached by one of four ways in (\ref{eq:add-rule}). Other
queues with $(N_{M},N_{B},N_{\bar{B}},N_{r},N_{\bar{r}})$ where $N_{r}\neq0$
or $N_{\bar{r}}\ne0$ can be reached by adding $q$ or $\bar{q}$
to the end of queues with smaller $N_{M}$, $N_{B}$ and $N_{\bar{B}}$.
Queues marked as $(N_{M},N_{B},N_{\bar{B}},1,0)$ can be reached by
\begin{align}
(a) & [(N_{M},N_{B},N_{\bar{B}},0,0)-(N_{M},N_{B},N_{\bar{B}}-1,0,3)]+q,\nonumber \\
(b) & (N_{M}-1,N_{B},N_{\bar{B}},2,0)+\bar{q}.\label{eq:gqcr-s1}
\end{align}
In manner $(a)$, the special queue $(N_{M},N_{B},N_{\bar{B}}-1,0,3)$
(which is included in $(N_{M},N_{B},N_{\bar{B}},0,0)$) is excluded
because of the step 4 in gQCR. Queues marked as $(N_{M},N_{B},N_{\bar{B}},0,1)$
can be reached by 
\begin{align}
(a) & [(N_{M},N_{B},N_{\bar{B}},0,0)-(N_{M},N_{B}-1,N_{\bar{B}},3,0)]+\bar{q},\nonumber \\
(b) & (N_{M}-1,N_{B},N_{\bar{B}},0,2)+q.\label{eq:gqcr-s2}
\end{align}
Queues marked as $(N_{M},N_{B},N_{\bar{B}},2,0)$ can be reached by
\begin{align}
(a) & (N_{M},N_{B},N_{\bar{B}},1,0)+q,\nonumber \\
(b) & (N_{M}-1,N_{B},N_{\bar{B}},3,0)+\bar{q}.\label{eq:gqcr-s3}
\end{align}
Queues marked as $(N_{M},N_{B},N_{\bar{B}},0,2)$ can be reached by
\begin{align}
(a) & (N_{M},N_{B},N_{\bar{B}},0,1)+\bar{q},\nonumber \\
(b) & (N_{M}-1,N_{B},N_{\bar{B}},0,3)+q.\label{eq:gqcr-s4}
\end{align}
The special queues $(N_{M},N_{B},N_{\bar{B}},3,0)$ and $(N_{M},N_{B},$
$N_{\bar{B}},0,3)$ can be build by normal ones 
\begin{align}
(N_{M},N_{B},N_{\bar{B}},3,0) & =(N_{M},N_{B},N_{\bar{B}},2,0)+q,\label{eq:gqcr-s5}\\
(N_{M},N_{B},N_{\bar{B}},0,3) & =(N_{M},N_{B},N_{\bar{B}},0,2)+\bar{q}.\label{eq:sqcr-s6}
\end{align}

Properties (\ref{eq:gqcr-s1}-\ref{eq:sqcr-s6}) lead to following
recursive equations, 
\begin{align}
F(N_{M},N_{B},N_{\bar{B}},1,0) & =F(N_{M},N_{B},N_{\bar{B}},0,0)\nonumber \\
 & -F(N_{M},N_{B},N_{\bar{B}}-1,0,2)\nonumber \\
 & +F(N_{M}-1,N_{B},N_{\bar{B}},2,0),\nonumber \\
F(N_{M},N_{B},N_{\bar{B}},0,1) & =F(N_{M},N_{B},N_{\bar{B}},0,0)\nonumber \\
 & -F(N_{M},N_{B}-1,N_{\bar{B}},2,0)\nonumber \\
 & +F(N_{M}-1,N_{B},N_{\bar{B}},0,2),\nonumber \\
F(N_{M},N_{B},N_{\bar{B}},2,0) & =F(N_{M},N_{B},N_{\bar{B}},1,0)\nonumber \\
 & +F(N_{M}-1,N_{B},N_{\bar{B}},2,0),\nonumber \\
F(N_{M},N_{B},N_{\bar{B}},0,2) & =F(N_{M},N_{B},N_{\bar{B}},0,1)\nonumber \\
 & +F(N_{M}-1,N_{B},N_{\bar{B}},0,2).\label{eq:gqcr-all}
\end{align}
For non-zero $N_{r}$ and $N_{\bar{r}}$, $F(N_{M},N_{B},N_{\bar{B}},N_{r},N_{\bar{r}})$
can be obtained with the help of two properties in Appendix \ref{sec:properties}.
Using them, we can derive the recursive equation for $F(N_{M},N_{B},$
$N_{\bar{B}},0,0)$ 
\begin{align}
 & F(N_{M},N_{B},N_{\bar{B}},0,0)\nonumber \\
 & =F(N_{M},N_{B}-1,N_{\bar{B}},0,0)+F(N_{M},N_{B},N_{\bar{B}}-1,0,0)\nonumber \\
 & -F(N_{M},N_{B}-1,N_{\bar{B}}-1,0,0)\nonumber \\
 & +6F(N_{M}-1,N_{B},N_{\bar{B}},0,0)\nonumber \\
 & -3F(N_{M}-1,N_{B}-1,N_{\bar{B}},0,0)\nonumber \\
 & -3F(N_{M}-1,N_{B},N_{\bar{B}}-1,0,0)\nonumber \\
 & -10F(N_{M}-2,N_{B},N_{\bar{B}},0,0)\nonumber \\
 & +F(N_{M}-2,N_{B}-1,N_{\bar{B}},0,0)\nonumber \\
 & +F(N_{M}-2,N_{B},N_{\bar{B}}-1,0,0)\nonumber \\
 & +4F(N_{M}-3,N_{B},N_{\bar{B}},0,0).\label{eq:recursive-f}
\end{align}
The detailed derivation of Eq. (\ref{eq:recursive-f}) is given in Appendix
\ref{sec:recursive-f}. 

\subsection{Solution to recursive equation by generating functions for gQCR}

We start with the most simple case $F(N_{M},0,0,0,0)$ with $N_{M}\geqslant1$,
Eq. (\ref{eq:recursive-f}) is simplified as 
\begin{align*}
 & F(N_{M},0,0,0,0)\\
 & =6F(N_{M}-1,0,0,0,0)-10F(N_{M}-2,0,0,0,0)\\
 & +4F(N_{M}-3,0,0,0,0).
\end{align*}
With the initial values $F(0,0,0,0,0)=1$, $F(1,0,0,0,0)=2$, $F(2,0,0,0,0)=6$,
and $F(3,0,0,0,0)=20$, we obtain the solution 
\begin{equation}
F(N_{M},0,0,0,0)=\frac{1}{2}\left[\left(2+\sqrt{2}\right)^{N_{M}}+\left(2-\sqrt{2}\right)^{N_{M}}\right].
\end{equation}

\end{multicols}
Now we multiply Eq. (\ref{eq:recursive-f}) by $x^{N_{M}}$ and sum
over $N_{M}\geqslant3$, we obtain the recursive equation Eq. (\ref{eq:rec-ax1})
for the partial generating function $A(x;N_{B},N_{\bar{B}})$ defined
in Eq. (\ref{eq:part_gen_func}). In the special case with $N_{B}=0$
in Eq. (\ref{eq:rec-ax1}) we obtain 
\begin{align}
A(x;0,N_{\bar{B}}) & =\frac{(1-3x+x^{2})^{N_{\bar{B}}}}{(1-6x+10x^{2}-4x^{3})^{N_{\bar{B}}}}\cdot\frac{1-2x}{1-4x+2x^{2}},\label{eq:ax-nbb}
\end{align}
where we have used 
\begin{equation}
A(x;0,0)  =  \frac{1-2x}{1-4x+2x^{2}}.
\end{equation}
In the same way, we derive $A(x;N_{B},0)$ whose result is given by
Eq. (\ref{eq:ax-nbb}) by the replacement $N_{\bar{B}}\rightarrow N_{B}$.
We also obtain two summation properties 
\begin{align}
    & \sum_{N_{B}=0}^{\infty}A(x;N_{B},0)y^{N_{B}}=\frac{1-2x}{1-4x+2x^{2}}\cdot\frac{1-6x+10x^{2}-4x^{3}}{1-6x+10x^{2}-4x^{3}-y(1-3x+x^{2})},\label{eq:axyp1}\\
 & \sum_{N_{\bar{B}}=0}^{\infty}A(x;0,N_{\bar{B}})z^{N_{\bar{B}}}=\frac{1-2x}{1-4x+2x^{2}}\cdot\frac{1-6x+10x^{2}-4x^{3}}{1-6x+10x^{2}-4x^{3}-z(1-3x+x^{2})}.\label{eq:axyz}
\end{align}

We multiply Eq. (\ref{eq:rec-ax1}) by $y^{N_{B}}z^{N_{\bar{B}}}$
and take sums over $N_{B}\geqslant1$ and $N_{\bar{B}}\geqslant1$,
then we can solve $A(x,y,z)$ as 
\begin{equation}
A(x,y,z)  =  \frac{1-4x+4x^{2}-yz}{1-6x+10x^{2}-4x^{3}-y+3xy-x^{2}y-z+3xz-x^{2}z+yz}.\label{eq:gen-func-axyz}
\end{equation}
The detailed derivation of (\ref{eq:gen-func-axyz}) is shown in (\ref{eq:rec-ax}-\ref{eq:recursive-a2}). 

We can extract $F(N_{M},N_{B},N_{\bar{B}},0,0)$ as the coefficient
of $x^{N_{M}}y^{N_{B}}z^{N_{\bar{B}}}$ in polynomial expansion of
$A(x,y,z)$. The result is a sum of four terms, 
\begin{equation}
F(N_{M},N_{B},N_{\bar{B}},0,0)  =  I_{1}+I_{2}+I_{3}+I_{4},\label{eq:re_gqcr}
\end{equation}
where these terms are given by Eq. (\ref{eq:re_gqcr_all}) in Appendix
\ref{sec:coefficient-gqcr-1}.
\begin{multicols}{2}

\section{Particle number distribution for baryons and mesons }

\label{sec:compara_QCR}In a system of $N_{q}$ quark and $N_{\bar{q}}$
antiquark, we obtain in Eq. (\ref{eq:prob-nm-nb-nbbar}) the probability
$P(N_{M},N_{B},N_{\bar{B}})$ to form $N_{M}$ mesons, $N_{B}$ baryons
and $N_{\bar{B}}$ antibaryons. A general form of the raw moments
of meson and baryon numbers is 
\begin{align}
\overline{N_{M}^{m}N_{B}^{n}N_{\bar{B}}^{k}} & =\sum_{N_{M}N_{B}N_{\bar{B}}}N_{M}^{m}N_{B}^{n}N_{\bar{B}}^{k}\,P\left(N_{M},N_{B},N_{\bar{B}}\right)\nonumber \\
 & \times\delta_{N_{M}+3N_{B},N_{q}}\delta_{N_{M}+3N_{\bar{B}},N_{\bar{q}}},
\end{align}
where we use an \emph{overline} to denote the average at fixed quark
and antiquark numbers. The central moments for baryons and mesons
are related by the quark number conservation 
\begin{align}
\overline{\delta N_{B}^{n}} & =\overline{\delta N_{\bar{B}}^{n}},\nonumber \\
\overline{\delta N_{M}^{n}} & =\left(-3\right)^{n}\overline{\delta N_{B}^{n}},
\end{align}
where $\delta N_{i}\equiv N_{i}-\overline{N_{i}}$ with $i=M,B,\bar{B}$. 

In Fig. \ref{fig:Bmoments}, we show the ratios of the cumulants for
antibaryons as functions of quark-antiquark asymmetry 
\begin{equation}
z=\frac{N_{q}-N_{\bar{q}}}{N_{q}+N_{\bar{q}}},
\end{equation}
with the original QCR and gQCR at two values of total quark number
$x=N_{q}+N_{\bar{q}}$. The cumulants for antibaryons are given by 

\begin{align}
C_{1} & \equiv C_{1}^{\bar{B}}=\overline{N_{\bar{B}}},\nonumber \\
C_{2} & =\overline{\delta N_{B}^{2}}=\overline{\delta N_{\bar{B}}^{2}},\nonumber \\
C_{3} & =\overline{\delta N_{B}^{3}}=\overline{\delta N_{\bar{B}}^{3}},\nonumber \\
C_{4} & =\overline{\delta N_{B}^{4}}-3C_{2}^{2}=\overline{\delta N_{\bar{B}}^{4}}-3C_{2}^{2}.
\end{align}
Note that the first order cumulant for baryons is related to that
for antibaryons by $C_{1}^{B}=C_{1}+zx/3$. 
     As $x$ is not small ($x\gtrsim 200$ ), $C_{1}/x$, $C_{2}/C_{1}$, $C_{3}/C_{2}$
    and $C_{4}/C_{2}$ are almost independent of $x$ (i.e.,~the system size) and are only mainly dependent on the quark-antiquark asymmetry $z$ (proportional to the net-baryon density). Therefore, we plot their $z$ dependence under the original and generalized QCR respectively in Fig. \ref{fig:Bmoments}. 
    Panel (a) shows antibaryons are less produced
with gQCR than that the QCR. The antibaryon production with the gQCR
is more suppressed than the QCR for larger $z$. We can parameterize
the antibaryon number \cite{Song:2013isa} as $\overline{N_{\bar{B}}}/x=\left(z/3\right)\left(1-z\right)^{a}/\left[\left(1+z\right)^{a}-\left(1-z\right)^{a}\right]$
with $a=3$ for the QCR and $a=5$ for the gQCR. We see that the cumulant
ratios $C_{2}/C_{1}$, $C_{3}/C_{2}$ and $C_{4}/C_{2}$ tend to unity
at large $z$. This is because the antiquark number is small at large
$z$ and the aggregation of quarks and antiquarks in phase space to
form baryons and antibaryons is more stochastic and follows the Poisson
distribution. Because antibaryons are less produced in the gQCR and
thus the feature of the Poisson distribution is more obvious, which
can be seen by that the cumulant ratios in the gQCR approaches unity
more quickly at large $z$ than in the QCR. At small $z$, $C_{3}/C_{2}$
and $C_{4}/C_{2}$ are small and approach the Gaussian distribution. 

\begin{center}
    \includegraphics[width=0.9\linewidth]{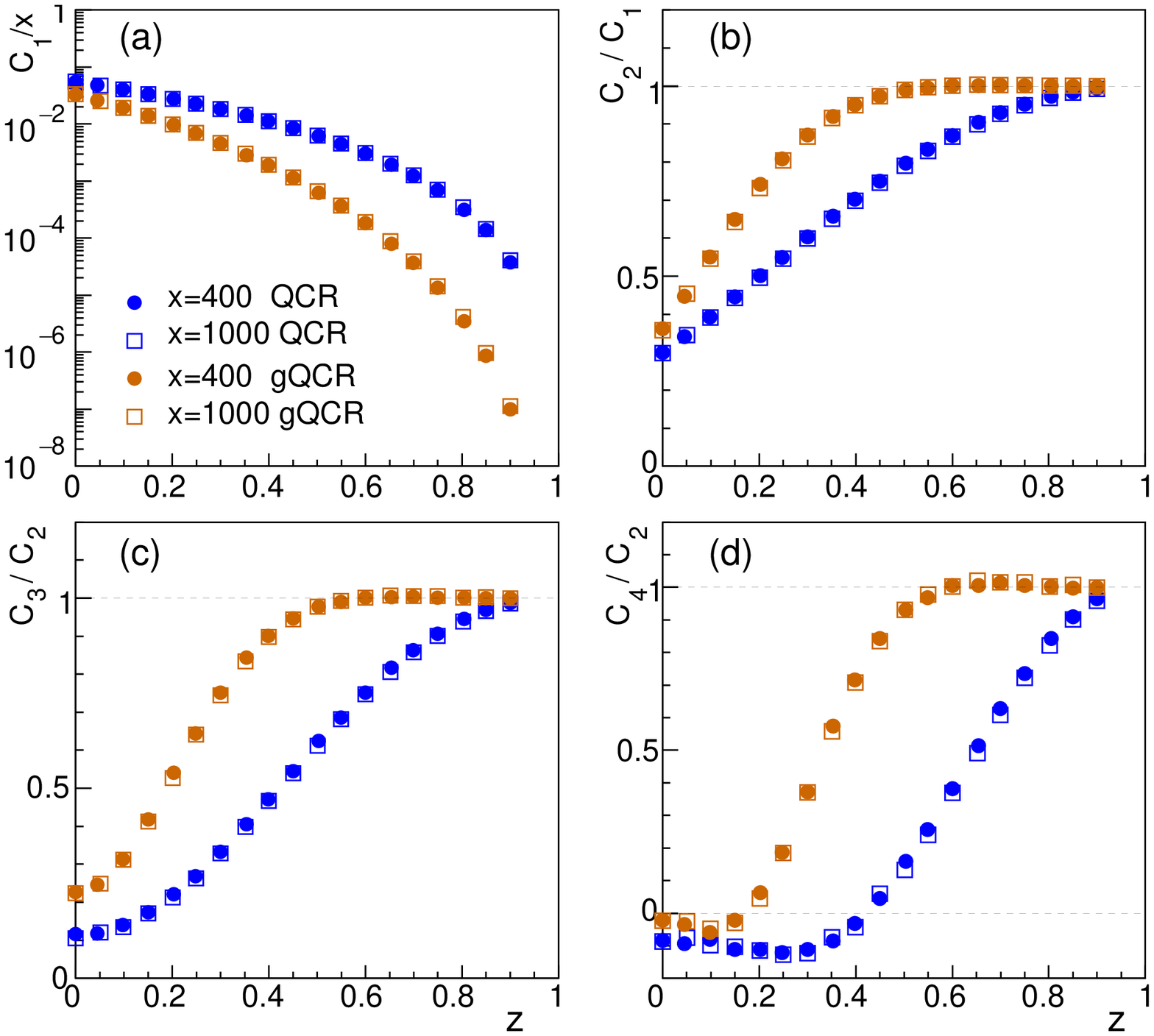}
    \figcaption{\label{fig:Bmoments} The ratios of cumulants for baryon or antibaryon number as functions
of quark-antiquark asymmetry $z$ with the original QCR and gQCR at
two different values of total quark number $x$: $C_{1}/x$, $C_{2}/C_{1}$,
$C_{3}/C_{2}$ and $C_{4}/C_{2}$.   }
\end{center}

\section{Multiplicity property of identified hadrons }

\label{sec:id_h_fc}Following the method of Ref. \cite{Song:2016ihg,Li:2017fhe},
we can obtain some multiplicity properties of identified hadrons by
taking advantage of the stochastic combination rule. In this paper,
we only consider the production of octet baryons with $J^{P}=(1/2)^{+}$
and decuplet baryons with $J^{P}=(3/2)^{+}$, pseudo-scalar mesons
with $J^{P}=0^{-}$ and vector mesons $J^{P}=1^{-}$ in the flavor
SU(3) ground state. The mean values of the multiplicities for identified
baryons and mesons are given by 

\begin{align}
\overline{N}_{B_{i}} & =g_{B_{i}}\frac{N_{B_{i}}^{\left(q\right)}}{N_{q}\left(N_{q}-1\right)\left(N_{q}-2\right)}\overline{N}_{B},\nonumber \\
\overline{N}_{M_{i}} & =g_{M_{i}}\frac{N_{M_{i}}^{\left(q\right)}}{N_{q}N_{\bar{q}}}\overline{N}_{M},\label{eq:av-nbi}
\end{align}
with 
\begin{align}
N_{B_{i}}^{\left(q\right)} & =S_{B_{i}}\prod_{f}\prod_{j=1}^{n_{f,B_{i}}}\left(N_{f}-j+1\right),\nonumber \\
N_{M_{i}}^{\left(q\right)} & =\prod_{f}\prod_{j=1}^{n_{f,M_{i}}}\left(N_{f}-j+1\right),
\end{align}
where $f$ denotes the quark or antiquark flavors in the baryon $B_{i}$
and the meson $M_{i}$, $n_{f,B_{i}}$ and $n_{f,M_{i}}$ are the
number of flavor $f$ quarks in $B_{i}$ and $M_{i}$ respectively,
$S_{B_{i}}$ counts the number of different permutations in quarks
and antiquarks, $g_{B_{i}}$ and $g_{M_{i}}$ are spin selection factors
for $B_{i}$ and $M_{i}$ respectively. We introduce a parameter $R_{V/P}$
to denote the relative weight of vector mesons to pseudo-scalar mesons
with the same quark content, and introduce parameter $R_{D/O}$ to
denote the relative weight of decuplet baryons to octet baryons with
the same quark content. We take $R_{V/P}=0.45$ and $R_{D/O}=0.5$
from studying the hadronic yields in relativistic high energy collisions \cite{Zhang:2012ff,Gou:2017foe}.
Because the values of the two parameters are extracted from the experimental data of hadronic yields, we emphasize that the two parameters can absorb the effects/contribution of various excited states and higher-mass resonances to a certain extent. 

Taking the proton as an example, we have $N_{\mathrm{p}}^{\left(q\right)}=3N_{u}\left(N_{u}-1\right)N_{d}$
for all possible combinations of $uud$. Then the ratio $N_{\mathrm{p}}^{\left(q\right)}/N_{q}\left(N_{q}-1\right)\left(N_{q}-2\right)$
stands for the formation probability of the proton having the quark
flavor $uud$. The spin selection factor for the proton is $g_{\mathrm{p}}=1/(1+R_{D/O})$.

We introduce the configuration of multiple hadrons as $\left\{ k_{\mathrm{h}_{1}}\mathrm{h}_{1},\,k_{\mathrm{h}_{2}}\mathrm{h}_{2},\,...\right\} \equiv\left\{ k_{\mathrm{h}}\mathrm{h}\right\} $,
where $\mathrm{h}_{i}$ denotes one hadron and $k_{\mathrm{h}_{i}}$
its number in the configuration. We define the joint moment of the
multiplicity for such a configuration as $N_{\left\{ k_{h}h\right\} }=\prod_{\mathrm{h}_{i}}N_{\mathrm{h}_{i}}^{\underline{k_{\mathrm{h}_{i}}}}$,
where we have used the falling factorial of the hadron multiplicity
$N_{\mathrm{h}}^{\underline{k_{\mathrm{h}}}}=\prod_{n=1}^{k_{\mathrm{h}}}\left(N_{\mathrm{h}}-n+1\right)$.
The average value of the joint moment of the multiplicity reads 
\begin{equation}
\overline{N_{\left\{ k_{\mathrm{h}}\mathrm{h}\right\} }}=\left(\prod_{\mathrm{h}_{i}}g_{\mathrm{h}_{i}}^{k_{\mathrm{h}_{i}}}\right)\frac{N_{\left\{ \mathrm{h}\right\} }^{\left(q\right)}}{N_{q}^{\underline{k_{q}}}N_{\bar{q}}^{\underline{k_{\bar{q}}}}}\overline{N_{B}^{\underline{k_{B}}}N_{\bar{B}}^{\underline{k_{\bar{B}}}}N_{M}^{\underline{k_{M}}},}\label{eq:gmhc}
\end{equation}
where $k_{B}=\sum_{\mathrm{h}_{i}}k_{\mathrm{h}_{i}}Q_{B,\mathrm{h}_{i}}$
counts the number of baryons in the multi-hadron configuration with
$Q_{B,\mathrm{h}_{i}}=1$ for $\mathrm{h}_{i}$ being a baryon and
$Q_{B,\mathrm{h}_{i}}=0$ for $\mathrm{h}_{i}$ being a meson and
an antibaryon, similarly $k_{\bar{B}}$ counts the number of antibaryons
and $k_{M}$ the number of mesons in the multi-hadron configuration,
$k_{q}=\sum_{\mathrm{h}_{i}}k_{\mathrm{h}_{i}}Q_{q,\mathrm{h}_{i}}$
counts the number of constituent quarks in the multi-hadron configuration
with $Q_{q,\mathrm{h}_{i}}=3$ for $\mathrm{h}_{i}$ being a baryon,
$Q_{q,\mathrm{h}_{i}}=0$ for $\mathrm{h}_{i}$ being an antibaryon
and $Q_{q,\mathrm{h}_{i}}=1$ for $\mathrm{h}_{i}$ being a meson,
similarly $k_{\bar{q}}$ counts the number of antiquarks. The numerator
in (\ref{eq:gmhc}) is defined as 
\begin{equation}
N_{\left\{ \mathrm{h}\right\} }^{\left(q\right)}=\left(\prod_{\mathrm{h}_{i}}S_{\mathrm{h}_{i}}\right)\prod_{f}\prod_{j=1}^{n_{f}}\left(N_{f}-j+1\right),
\end{equation}
denotes the number of all possible combinations out of all quarks
with specific flavors in the hadrons in the multi-hadron configuration,
where $f$ runs over $u,d,s,\bar{u},\bar{d},\bar{s}$ and $n_{f}=\sum_{\mathrm{h}_{i}}k_{\mathrm{h}_{i}}n_{f,\mathrm{h}_{i}}$
counts the number of $f$ flavor quarks or antiquarks in the multi-hadron
configuration. 

We take a few examples of multi-hadron configurations to illustrate
Eq. (\ref{eq:gmhc}). The first example is the configuration with
two protons, we have $k_{\mathrm{p}}=2$ and $N_{\mathrm{p}}^{\underline{2}}\equiv N_{\mathrm{p}}\left(N_{\mathrm{p}}-1\right)$,
so we have 
\begin{equation}
\overline{N_{\mathrm{p}}^{\underline{2}}}=\overline{N_{\mathrm{p}}\left(N_{\mathrm{p}}-1\right)}=g_{\mathrm{p}}^{2}\frac{N_{\mathrm{pp}}^{\left(q\right)}}{N_{q}^{\underline{6}}}\overline{N_{B}\left(N_{B}-1\right)},
\end{equation}
where $\overline{N_{B}\left(N_{B}-1\right)}$ denotes the average
for all possible baryon pairs, $N_{\mathrm{pp}}^{\left(q\right)}=3^{2}N_{u}^{\underline{4}}N_{d}^{\underline{2}}$
is the number of all possible combinations out of six quarks with
specific flavors in two protons. The ratio $N_{\mathrm{pp}}^{\left(q\right)}/N_{q}^{\underline{6}}$
gives the probability of the six-quark combination having the particular
flavor structure $(uud)(uud)$. The second example is the configuration
$\mathrm{pp\bar{n}}$, we have 
\begin{align}
\overline{N_{\mathrm{p}}^{\underline{2}}N_{\bar{\mathrm{n}}}} & =\overline{N_{\mathrm{p}}\left(N_{\mathrm{p}}-1\right)N_{\bar{\mathrm{n}}}}\nonumber \\
 & =\left(g_{\mathrm{p}}^{2}\frac{N_{\mathrm{pp}}^{\left(q\right)}}{N_{q}^{\underline{6}}}\right)\left(g_{\bar{\mathrm{n}}}\frac{N_{\bar{\mathrm{n}}}^{\left(q\right)}}{N_{\bar{q}}^{\underline{3}}}\right)\overline{N_{B}\left(N_{B}-1\right)N_{\bar{B}}},
\end{align}
where the first parentheses in the second line gives the probability
of two baryons having the flavor and spin structure of two protons
and the second parentheses gives the probability of an antibaryon
having the flavor and spin structure of the antineutron. 

The all orders of moments and correlation functions of hadron multiplicity
can be built from Eq. (\ref{eq:gmhc}). The two-body correlation function
reads 
\begin{align}
C_{\alpha\beta} & \equiv\overline{\delta N_{\alpha}\delta N_{\beta}}\nonumber \\
 & =\overline{N_{\alpha}N_{\beta}}-\overline{N}_{\alpha}\overline{N}_{\beta}=\overline{N}_{\alpha\beta}+\delta_{\alpha,\beta}\overline{N}_{\alpha}-\overline{N}_{\alpha}\overline{N}_{\beta},
\end{align}
where $\alpha$, $\beta$ denote two hadrons, and we have used $\overline{N}_{\alpha\beta}\equiv\overline{N_{\alpha}N_{\beta}}$
for $\alpha\neq\beta$ and $\overline{N}_{\alpha\beta}\equiv\overline{N_{\alpha}^{\underline{2}}}$
for $\alpha=\beta$. The three-body correlation function reads 
\begin{align}
C_{\alpha\beta\gamma} & \equiv\overline{\delta N_{\alpha}\delta N_{\beta}\delta N_{\gamma}}\nonumber \\
 & =\overline{N_{\alpha}N_{\beta}N_{\gamma}}-\overline{N}_{\alpha}C_{\beta\gamma}-\overline{N}_{\beta} C_{\alpha\gamma}  -\overline{N}_{\gamma}C_{\alpha\beta}   \nonumber \\
 & -\overline{N}_{\alpha}\overline{N}_{\beta}\overline{N}_{\gamma},
\end{align}
where $\overline{N_{\alpha}N_{\beta}N_{\gamma}}$ can be expressed
by falling factorials, 
\begin{align}
\overline{N_{\alpha}N_{\beta}N_{\gamma}} & =\overline{N}_{\alpha\beta\gamma}+\delta_{\alpha,\beta}\overline{N}_{\alpha\gamma}+\delta_{\alpha,\gamma}\overline{N}_{\alpha\beta}+\delta_{\beta,\gamma}\overline{N}_{\alpha\beta} \nonumber \\
 & +\delta_{\alpha,\beta}\delta_{\alpha,\gamma}\overline{N}_{\alpha}.
\end{align}
Here we have used $\overline{N}_{\alpha\beta\gamma}\equiv\overline{N_{\alpha}N_{\beta}N_{\gamma}}$
for $\alpha\neq\beta\neq\gamma$, $\overline{N}_{\alpha\beta\gamma}\equiv\overline{N_{\alpha}^{\underline{2}}N_{\gamma}}$
for $\alpha=\beta\neq\gamma$ and $\overline{N}_{\alpha\beta\gamma}\equiv\overline{N_{\alpha}^{\underline{3}}}$
for $\alpha=\beta=\gamma$. Note that $\overline{N}_{\alpha\beta\gamma}$
is symmetric for any permutation of $\alpha$, $\beta$ and $\gamma$.
The four-body correlation function can be written as 
\begin{align}
C_{\alpha\beta\gamma\epsilon} & =\overline{\delta N_{\alpha}\delta N_{\beta}\delta N_{\gamma}\delta N_{\epsilon}}\nonumber \\
 & =\overline{N_{\alpha}N_{\beta}N_{\gamma}N_{\epsilon}} -\left(\overline{N}_{\alpha}C_{\beta\gamma\epsilon}+\mathrm{permutation}\right) \\
 & -\left(\overline{N}_{\alpha}\overline{N}_{\beta}C_{\gamma\epsilon}+\mathrm{permutation}\right) -\overline{N}_{\alpha}\overline{N}_{\beta}\overline{N}_{\gamma}\overline{N}_{\epsilon},\nonumber
\end{align}
where $\overline{N_{\alpha}N_{\beta}N_{\gamma}N_{\epsilon}}$ can be expressed by falling factorials 
\begin{align}
&\overline{N_{\alpha}N_{\beta}N_{\gamma}N_{\epsilon}} \nonumber \\
& =\overline{N}_{\alpha\beta\gamma\epsilon} +\left(\delta_{\alpha,\beta}\overline{N}_{\alpha\gamma\epsilon}+\mathrm{permutation}\right) \\
 & +\left(\delta_{\alpha,\gamma}\delta_{\beta,\epsilon}\overline{N}_{\alpha\beta}+\mathrm{permutation}\right)+\delta_{\alpha,\beta}\delta_{\alpha,\gamma}\delta_{\alpha,\epsilon}\overline{N}_{\alpha}.\nonumber
\end{align}
Here we have used $\overline{N}_{\alpha\beta\gamma\epsilon}\equiv\overline{N_{\alpha}N_{\beta}N_{\gamma}N_{\epsilon}}$
for $\alpha\neq\beta\neq\gamma\neq\epsilon$, $\overline{N}_{\alpha\beta\gamma\epsilon}\equiv\overline{N_{\alpha}^{\underline{2}}N_{\gamma}N_{\epsilon}}$
for $\alpha=\beta\neq\gamma\neq\epsilon$, $\overline{N}_{\alpha\beta\gamma\epsilon}\equiv\overline{N_{\alpha}^{\underline{3}}N_{\epsilon}}$
for $\alpha=\beta=\gamma\neq\epsilon$ and $\overline{N}_{\alpha\beta\gamma\epsilon}\equiv\overline{N_{\alpha}^{\underline{4}}}$
for $\alpha=\beta=\gamma=\epsilon$. Note that $\overline{N}_{\alpha\beta\gamma\epsilon}$
is symmetric for any permutation of $\alpha$, $\beta$, $\gamma$
and $\epsilon$. 

The cumulants of the net proton number $N_{\mathrm{p}}-N_{\bar{\mathrm{p}}}$
can be calculated by combinations of above multi-body correlation
functions (except $C_{1}$ which is just the mean value of the net proton number)
\begin{align}
C_{1} & =\overline{N}_{\mathrm{p}}-\overline{N}_{\bar{\mathrm{p}}},\nonumber \\
C_{2} & =C_{\mathrm{pp}}-2C_{\mathrm{p}\bar{\mathrm{p}}}+C_{\bar{\mathrm{p}}\bar{\mathrm{p}}},\nonumber \\
C_{3} & =C_{\mathrm{ppp}}-3C_{\mathrm{pp}\bar{\mathrm{p}}}+3C_{\mathrm{p}\bar{\mathrm{p}}\bar{\mathrm{p}}}-C_{\bar{\mathrm{p}}\bar{\mathrm{p}}\bar{\mathrm{p}}},\nonumber \\
C_{4} & =C_{\mathrm{pppp}}-4C_{\mathrm{ppp}\bar{\mathrm{p}}}+6C_{\mathrm{pp}\bar{\mathrm{p}}\bar{\mathrm{p}}} -4C_{\mathrm{p}\bar{\mathrm{p}}\bar{\mathrm{p}}\bar{\mathrm{p}}}+C_{\bar{\mathrm{p}}\bar{\mathrm{p}}\bar{\mathrm{p}}\bar{\mathrm{p}}}-3C_{2}^{2}.
\end{align}
In Fig. \ref{fig:netProtonCn}, we show the cumulant ratios for the
net proton number with the QCR and the gQCR at given total quark number
$x=2000$ as functions of the quark-antiquark asymmetry $z$. We have
checked that the cumulant ratios of the net proton number are independent
of $x$ for large $x$. Here we assume the number of strange quarks
is $N_{s}=N_{\bar{s}}=0.45N_{\bar{u}}$, where the strangeness conservation
is satisfied and strangeness suppression factor 0.45 is consistent with the observation in relativistic heavy-ion collisions \cite{Shao:2009uk}. Because the net-baryon number is fixed at given numbers
of quarks and antiquarks, Fig. \ref{fig:netProtonCn} only show the fluctuations of the net proton number brought by the quark combination process. 
$C_{1}/C_{2}$ and $C_{3}/C_{2}$ of net proton number increase with $z$ and approach to 1.5 and $1/3$ as $z\to1$, respectively. $C_{4}/C_{2}$ decreases with $z$ and approaches to $-1/3$ as $z\to1$. These results are different from those of the statistical model for hadron resonance gas with thermal equilibrium \cite{Nahrgang:2014fza}, in which $C_{1}/C_{2}$ and $C_{3}/C_{2}$ increase to one and $C_{4}/C_{2}$ almost keeps a constant of one at large baryon number chemical potential.

\begin{center}
    \includegraphics[width=0.9\linewidth]{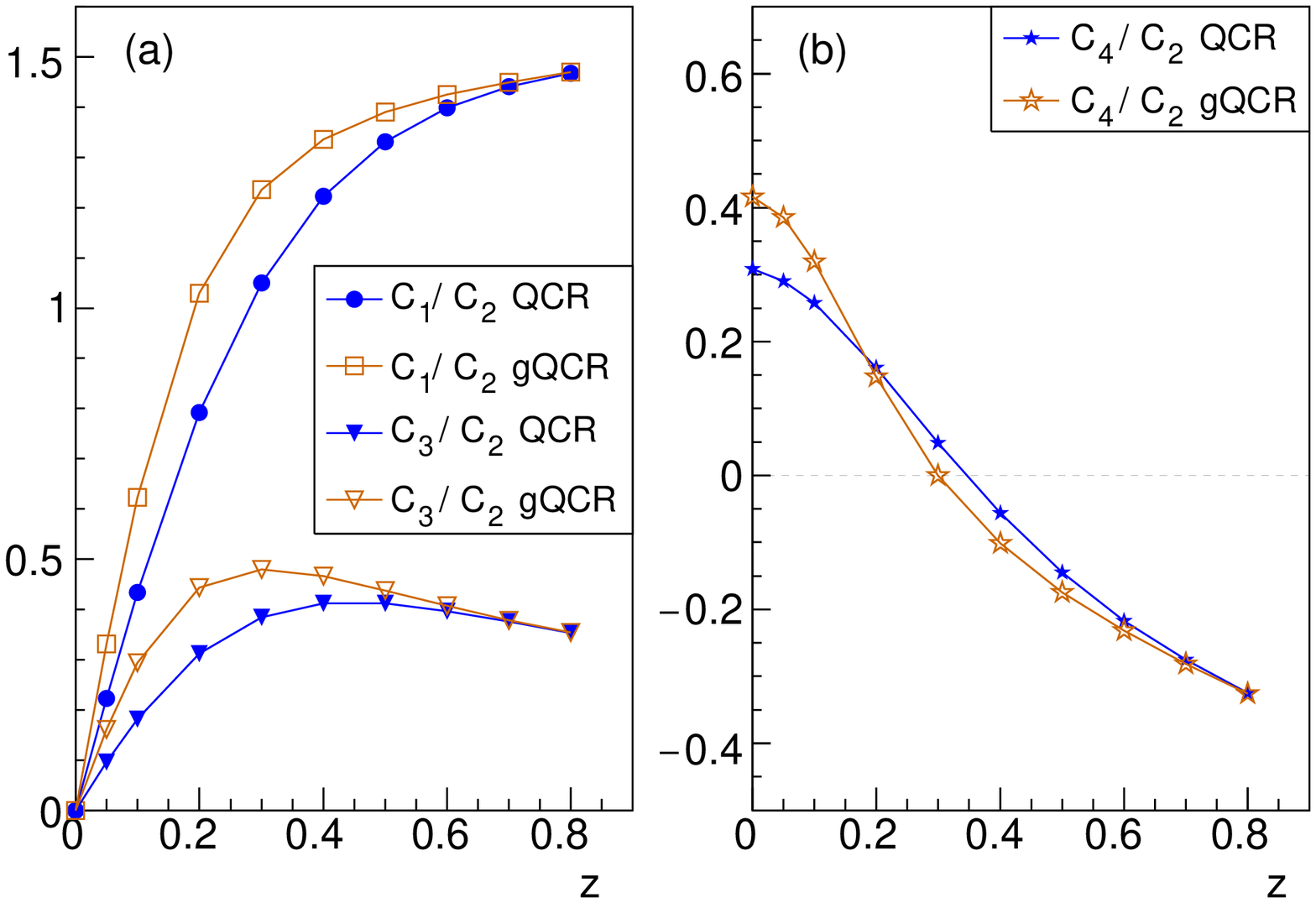}
    \figcaption{\label{fig:netProtonCn} The cumulant ratios of the net proton number directly-produced through the QCR and the gQCR as functions of quark-antiquark asymmetry $z$ at $x=2000$.  }
\end{center}

\section{Effects of quark number fluctuations at hadronization\label{sec:incl_qnfc}}

The numbers of quarks and antiquarks may fluctuate in heavy energy
collisions in event by event, so hadronic observables should be influenced
by the fluctuations at the quark level. We denote the distribution
of quark numbers at hadronization as 
\begin{equation}
P\left(\{N_{f}\}\right)\equiv P\left(N_{u},N_{\bar{u}},N_{d},N_{\bar{d}},N_{s},N_{\bar{s}}\right),
\end{equation}
and the event average of a hadronic observable $A_{\mathrm{h}}$ is given by 
\begin{align}
\langle A_{\mathrm{h}}\rangle & =\sum_{\{N_{f}\}}P\left(\{N_{f}\}\right)\overline{A}_{\mathrm{h}}\left(\{N_{f}\}\right),\label{eq:ahave}
\end{align}
 where $\overline{A}_{\mathrm{h}}\left(\{N_{f}\}\right)$ denote the
average at fixed quark and antiquark numbers which is obtained in
the previous section. 

In practical evaluation, it is more convenient to take the expansion
of $\overline{A}_{\mathrm{h}}\left(\{N_{f}\}\right)$ in $N_{f}$
around the event-average numbers of quarks and antiquarks $\{\langle N_{f}\rangle\}$
as 
\begin{align}
\overline{A}_{\mathrm{h}} & =\overline{A}_{\mathrm{h},0}+\sum_{f}\left.\frac{\partial\overline{A}_{\mathrm{h}}}{\partial N_{f}}\right|_{0}\delta N_{f}\nonumber \\
 & +\frac{1}{2}\sum_{f_{1}f_{2}}\left.\frac{\partial^{2}\overline{A}_{\mathrm{h}}}{\partial N_{f_{1}}\partial N_{f_{2}}}\right|_{0}\delta N_{f_{1}}\delta N_{f_{2}} +\cdots,
\end{align}
where $\delta N_{f}=N_{f}-\langle N_{f}\rangle$ and the subscript
'0' denotes the values at $\langle N_{f}\rangle$. Substituting it
into Eq. (\ref{eq:ahave}), we obtain 
\begin{align}
\langle A_{\mathrm{h}}\rangle & =\overline{A}_{\mathrm{h},0}+\frac{1}{2}\sum_{f_{1}f_{2}}\left.\frac{\partial^{2}\overline{A}_{\mathrm{h}}}{\partial N_{f_{1}}\partial N_{f_{2}}}\right|_{0}\left\langle \delta N_{f_{1}}\delta N_{f_{2}}\right\rangle \nonumber \\
 & +\frac{1}{3!}\sum_{f_{1}f_{2}f_{3}}\left.\frac{\partial^{3}\overline{A}_{\mathrm{h}}}{\partial N_{f_{1}}\partial N_{f_{2}}\partial N_{f_{3}}}\right|_{0}\left\langle \delta N_{f_{1}}\delta N_{f_{2}}\delta N_{f_{3}}\right\rangle \nonumber \\
 & +....,\label{eq:average}
\end{align}
which involves multi-body correlations for the quark and antiquark
number with the quark number distribution $P\left(\{N_{f}\}\right)$
\begin{align}
C_{f_{1}f_{2}} & \equiv\left\langle \delta N_{f_{1}}\delta N_{f_{2}}\right\rangle ,\nonumber \\
C_{f_{1}f_{2}f_{3}} & \equiv\left\langle \delta N_{f_{1}}\delta N_{f_{2}}\delta N_{f_{3}}\right\rangle ,\nonumber \\
C_{f_{1}f_{2}f_{3}f_{4}} & \equiv\left\langle \delta N_{f_{1}}\delta N_{f_{2}}\delta N_{f_{3}}\delta N_{f_{4}}\right\rangle ,\nonumber \\
\cdots & \cdots
\end{align}
where we have used the same symbols $C$ as in Sec. \ref{sec:id_h_fc} to denote the quark number correlation functions but with quark flavor indices. 

Using the above moment expansion method, we can study, for the selected phase-space window such as midrapidity region $|y|<0.5$, the influence of different quark and antiquark distributions in the window on the production of hadrons at hadronization. This extends the applicability of QCR described in Sec.~\ref{sec:old_QCR} and \ref{sec:new_QCR} where only the stochastic quark-antiquark population is considered.

\section{Influence of resonance decays}

\label{sec:decay_formulas}The long life (stable) hadrons measured
in experiments include contributions from resonance decays. In this
section we consider the influence of resonance decays to the multiplicity
correlations of long life hadrons. For a resonance $\mathrm{h}_{i}$,
its stable daughters are denoted as $\alpha$, $\beta$, $\gamma$,
etc., and the corresponding decay branching ratios are $D_{i\alpha}$,
$D_{i\beta}$, $D_{i\gamma}$, etc., respectively. The distribution
function of $\left\{ N_{\alpha},N_{\beta},N_{\gamma},\cdots\right\} $
from decays of the resonance $\mathrm{h}_{i}$ with the particle number
$N_{\mathrm{h}_{i}}$ is denoted as $f\left(N_{\mathrm{h}_{i}},\left\{ N_{\alpha}^{i}\right\} \right)$
where $N_{\alpha}^{i}$ denotes number of the stable hadron $\alpha$
from the resonance $\mathrm{h}_{i}$. We take the multi-nominal distribution
for selection of decay channels. Convoluting with the joint distribution
of directly produced hadrons $P\left(\left\{ N_{\mathrm{h}_{i}}\right\} ,\left\{ \left\langle N_{f}\right\rangle \right\} \right)$
at hadronization, we get the joint distributions of stable hadrons
$F\left(\left\{ N_{\alpha},N_{\beta},N_{\gamma},\cdots\right\} \right)$
\cite{Song:2016ihg}. 

From the joint distribution functions of stable hadrons, we get the
average yield of a stable hadron $\alpha$ 
\begin{align}
\langle N_{\alpha}\rangle & =\sum_{i}\langle N_{i}\rangle D_{i\alpha},\label{eq:final_nhi}
\end{align}
where the index $i$ runs over all directly produced hadron species
including stable hadrons (we define $D_{\alpha\alpha}=1$), and we
have used shorthand notation $N_{i}\equiv N_{\mathrm{h}_{i}}$. The
two-body multiplicity correlation function is 
\begin{align}
C_{\alpha\beta} & =\sum_{ij}C_{ij}D_{i\alpha}D_{j\beta}+\sum_{i}\langle N_{i}\rangle D_{i\alpha}\left(\delta_{\alpha\beta}-D_{i\beta}\right).
\end{align}
The three-body multiplicity correlation function of stable hadrons is 
\begin{align}
C_{\alpha\beta\gamma} & =\sum_{ijk}C_{ijk}D_{i\alpha}D_{j\beta}D_{k\gamma} +\sum_{ij}C_{ij}D_{i\alpha}\left[\delta_{\alpha\beta}-D_{i\beta}\right]D_{j\gamma}\nonumber \\
 & +\sum_{ij}C_{ij}D_{i\alpha}D_{j\beta}\left\{ \left[\delta_{\alpha\gamma}-D_{i\gamma}\right]+\left[\delta_{\beta\gamma}-D_{j\gamma}\right]\right\} \nonumber \\
 & +\sum_{i}\langle N_{i}\rangle D_{i\alpha}\left[\left(\delta_{\alpha\beta}-D_{i\beta}\right)\left(\delta_{\alpha\gamma}-D_{i\gamma}\right)\right.\nonumber \\
 & \left.+D_{i\beta}\left(D_{i\gamma}-\delta_{\beta\gamma}\right)\right].
\end{align}
The four-body multiplicity correlation function of stable hadrons is
\end{multicols}
\begin{align}
C_{\alpha\beta\gamma\epsilon} & =\sum_{ijkl}C_{ijkl}D_{i\alpha}D_{j\beta}D_{k\gamma}D_{l\epsilon}\nonumber \\
 & +\sum_{ijk}\left[C_{ijk}+\langle N_{i}\rangle C_{jk}\right]\biggl\{\left(\delta_{\alpha\beta}-D_{i\beta}\right)D_{i\alpha}D_{j\gamma}D_{k\epsilon}+\left(\delta_{\alpha\gamma}-D_{i\gamma}\right)D_{i\alpha}D_{j\beta}D_{k\epsilon}\nonumber \\
 & +\left(\delta_{\alpha\epsilon}-D_{i\epsilon}\right)D_{i\alpha}D_{j\gamma}D_{k\beta}+\left(\delta_{\beta\gamma}-D_{i\gamma}\right)D_{i\beta}D_{j\alpha}D_{k\epsilon}+\left(\delta_{\beta\epsilon}-D_{i\epsilon}\right)D_{i\beta}D_{j\alpha}D_{k\gamma} +\left(\delta_{\gamma\epsilon}-D_{i\epsilon}\right)D_{i\gamma}D_{j\alpha}D_{k\beta}\biggr\}\nonumber \\
 & +\sum_{ij}\left[C_{ij}+\langle N_{i}\rangle\langle N_{j}\rangle\right]\biggl\{ D_{i\alpha}\left(\delta_{\alpha\beta}-D_{i\beta}\right)D_{j\gamma}\left(\delta_{\gamma\epsilon}-D_{j\epsilon}\right)\nonumber \\
 & +D_{i\alpha}\left(\delta_{\alpha\gamma}-D_{i\gamma}\right)D_{j\beta}\left(\delta_{\beta\epsilon}-D_{j\epsilon}\right)+D_{i\alpha}\left(\delta_{\alpha\epsilon}-D_{i\epsilon}\right)D_{j\beta}\left(\delta_{\beta\gamma}-D_{j\gamma}\right)\biggr\}\nonumber \\
 & +\sum_{ij}C_{ij}D_{i\alpha}\left\{ \left(\delta_{\alpha\beta}-D_{i\beta}\right)\left(\delta_{\alpha\gamma}-D_{i\gamma}\right)+D_{i\beta}\left(D_{i\gamma}-\delta_{\beta\gamma}\right)\right\} D_{j\epsilon} +\sum_{ij}C_{ij}D_{i\alpha}\left\{ \left(\delta_{\alpha\beta}-D_{i\beta}\right)\left(\delta_{\alpha\epsilon}-D_{i\epsilon}\right)+D_{i\beta}\left(D_{i\epsilon}-\delta_{\beta\epsilon}\right)\right\} D_{j\gamma}\nonumber \\
 & +\sum_{ij}C_{ij}D_{i\alpha}\left\{ \left(\delta_{\alpha\gamma}-D_{i\gamma}\right)\left(\delta_{\alpha\epsilon}-D_{i\epsilon}\right)+D_{i\gamma}\left(D_{i\epsilon}-\delta_{\gamma\epsilon}\right)\right\} D_{j\beta}+\sum_{ij}C_{ij}D_{i\beta}\left\{ \left(\delta_{\beta\gamma}-D_{i\gamma}\right)\left(\delta_{\beta\epsilon}-D_{i\epsilon}\right)+D_{i\gamma}\left(D_{i\epsilon}-\delta_{\gamma\epsilon}\right)\right\} D_{j\alpha}\nonumber \\
 & +\sum_{i}\langle N_{i}\rangle D_{i\alpha}\biggl\{-6D_{i\beta}D_{i\gamma}D_{i\epsilon} +2\left[\delta_{\alpha\beta}D_{i\gamma}D_{i\epsilon}+\left(\delta_{\alpha\gamma}+\delta_{\beta\gamma}\right)D_{i\beta}D_{i\epsilon}+\left(\delta_{\alpha\epsilon}+\delta_{\beta\epsilon}+\delta_{\gamma\epsilon}\right)D_{i\beta}D_{i\gamma}\right]\nonumber \\
 & -\left[\left(\delta_{\alpha\gamma}\delta_{\beta\epsilon}+\delta_{\alpha\epsilon}\delta_{\beta\gamma}+\delta_{\alpha\gamma}\delta_{\alpha\epsilon}+\delta_{\beta\gamma}\delta_{\beta\epsilon}\right)D_{i\beta}\right. \left.+\left(\delta_{\alpha\beta}\delta_{\gamma\epsilon}+\delta_{\alpha\beta}\delta_{\alpha\epsilon}\right)D_{i\gamma}+\delta_{\alpha\beta}\delta_{\alpha\gamma}D_{i\epsilon}\right]+\delta_{\alpha\beta}\delta_{\alpha\gamma}\delta_{\alpha\epsilon}\biggl\}.
\end{align}
\begin{multicols}{2}
The higher order multiplicity correlation functions can similarly
be derived from the joint distribution functions of stable hadrons. 

\section{Application: cumulants for net protons in heavy ion collisions }

\label{sec:Results-and-discussions}In this section, we take a simple
example of a quark system and calculate the cumulant ratios for net
protons in the final state with gQCR. 
We will discuss our results in the context of the experimental observation in Au+Au collisions at RHIC. 

We consider a quark system having the property $S\sigma\equiv C_{3}/C_{2}=z$
and $\kappa\sigma^{2}\equiv C_{4}/C_{2}=1$ for the baryon quantum
number, which is similar to the base line in the grand canonical ensemble
in statistical model \cite{Karsch:2010ck}. A simple case for the
quark number correlation functions satisfying the above property is
\begin{align}
C_{ff} & =\langle N_{f}\rangle,\nonumber \\
C_{fff} & =3\langle N_{f}\rangle,\nonumber \\
C_{ffff} & =9\langle N_{f}\rangle+3\langle N_{f}\rangle^{2},
\end{align}
for diagonal elements and non-vanishing off-diagonal elements ($f_{1}\neq f_{2}$)
\begin{equation}
C_{f_{1}f_{1}f_{2}f_{2}}=\langle N_{f_{1}}\rangle\langle N_{f_{2}}\rangle.
\end{equation}
We also assume following properties for correlation functions of the
strangeness as a result of strangeness conservation 
\begin{align}
C_{s\bar{s}} & =C_{ss},\nonumber \\
C_{ss\bar{s}} & =C_{s\bar{s}\bar{s}}=C_{sss},\nonumber \\
C_{ss\bar{s}\bar{s}} & =C_{ssss}.
\end{align}

    Because total quark number $x$ in midrapidity region $|y|<0.5$ in central heavy-ion collisions at RHIC energy is usually large ($x\gtrsim$500), the cumulant ratios for net protons in the final state in our model is not sensitive  to $x$ (the system size). Here, we just take $x=5000$ in the calculation. We use Eq. (\ref{eq:final_nhi}) to fit the
data of the $\bar{\mathrm{p}}/\mathrm{p}$ yield ratio in Au+Au collisions and determine $z$ in the collision energy range \textbf{$\sqrt{s_{NN}}\in[7,200]$} GeV. 

In Fig. \ref{fig:cumulant_net_p_pt1}, we show the results for the
cumulant ratios for net protons in the final state at different collisional
energies.  Our results only incorporate
the contributions of quark number fluctuations and correlations up
to fourth order as in Eq. (\ref{eq:average}). The auxiliary horizontal
axis shows the corresponding quark-antiquark asymmetry parameter $z$.
Results including the contributions from strong and electromagnetic
(S\&EM) decays of resonances are in dashed lines, while results with
full decay contributions including weak decays are in solid lines.
The STAR data are shown in solid circles with error bars.

\begin{center}
\includegraphics[width=0.9\linewidth]{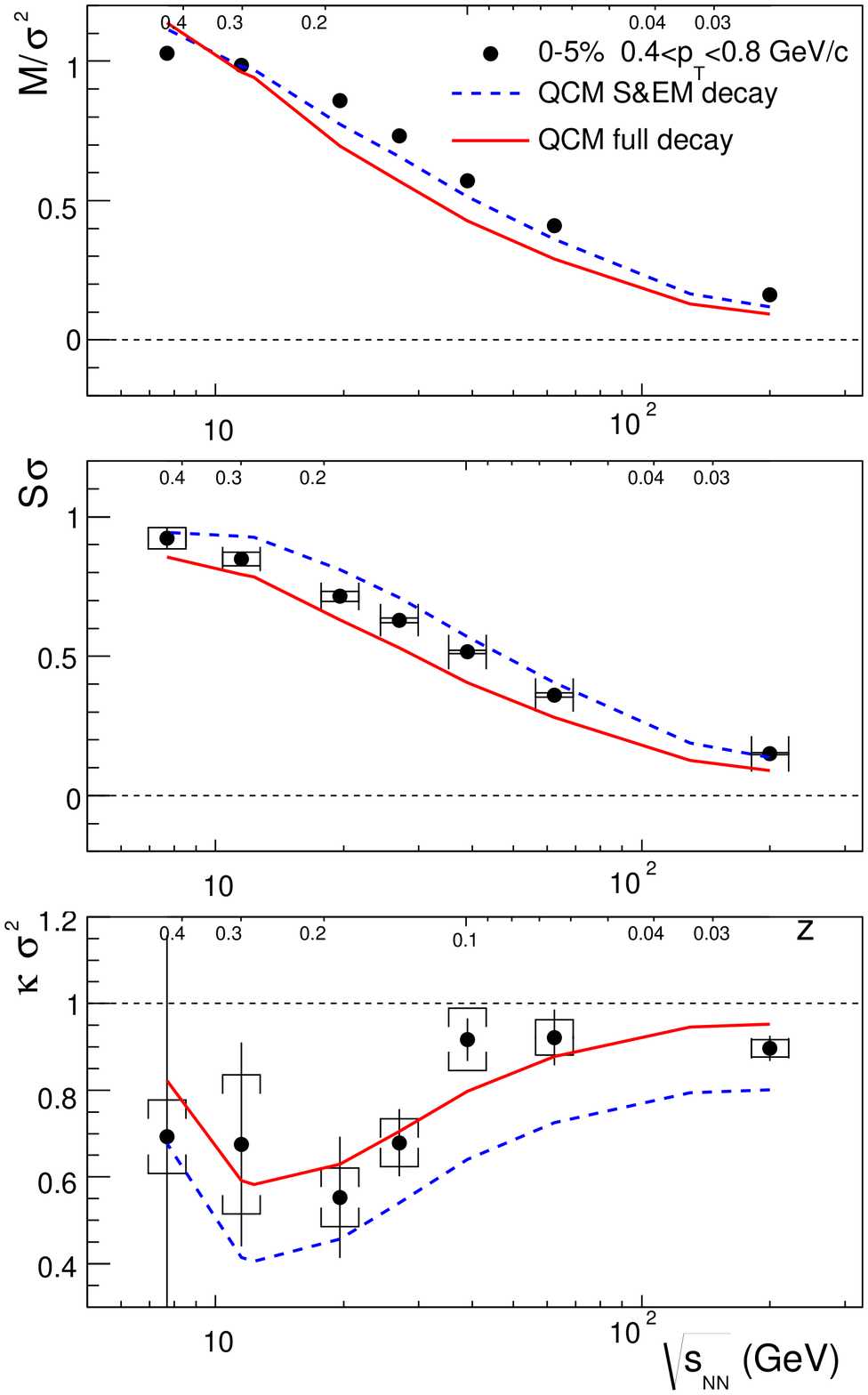}
    \figcaption{\label{fig:cumulant_net_p_pt1} The cumulant ratios for net proton at different collision energies.
Solid circles with error bars are experimental data \cite{Adamczyk:2013dal,Luo:2015ewa}.
Solid and dashed lines are theoretical results of SDQCM. }
\end{center}

We see in the figure that the cumulant ratios for net protons as functions
of collisional energies in our model can describe the experimental
data: $M/\sigma^{2}$ and $S\sigma$ increase with $z$ or equivalently
decrease with collisional energies, while $\kappa\sigma^{2}$ decreases
with collisional energies until it reaches a minimum at $\sqrt{s_{NN}}=$20
GeV and then increases toward unity at high energies. Our results
for $\kappa\sigma^{2}$ are the consequence of the competition between
two effects: (1) The cumulant ratio $C_{4}/C_{2}$ from directly produced
net protons by quark combination, as shown in Fig. \ref{fig:netProtonCn},
always decreases with increasing $z$; (2) The cumulant ratio $C_{4}/C_{2}$
for baryons or antibaryons, as shown in Fig. \ref{fig:Bmoments},
rapidly increases as $z\gtrsim0.2$ (corresponding to $\sqrt{s_{NN}}\lesssim20$
GeV).

We now make some remarks about the results for $\kappa\sigma^{2}$
of net protons. Although our results for net protons can reproduce
the nontrivial dependence on collisional energies, we always have
$\kappa\sigma^{2}=1$ for net baryons at all collisional energies
due to the flavor/charge conservation in quark combination. Therefore,
our results indicate that cumulant ratios of net proton number in this simple case do not exactly follow those of the net baryon number. 
 This is different from that of the statistical model \cite{Nahrgang:2014fza} for hadron resonance gas with thermal equilibrium which predicts the quite similar cumulant ratios for net-protons and net-baryons. 

    We emphasize that the current results are preliminary and are mainly used to show the potential application of our model in hadronic fluctuations and the related phase transition in relativistic heavy-ion collisions.
There are some limitations in the current calculations that should be clarified.
In this paper, we only consider the production of ground-state hadrons.  
Effects of higher-mass resonances are only partially absorbed by parameters $R_{V/P}$ and $R_{D/O}$.
According to our previous study \cite{Song:2016ihg}, the production of hadrons with small yields tends to follow Poisson distribution. Including those higher-mass resonances with small yields may enlarge the fourth moments of protons to a certain extent.  
Our model is a static one and does not consider the diffusion of hadrons/quarks during the finite hadronization time, which will cause some influence for the calculation of net-proton fluctuations in a specific window. 
In addition, in order to make final comparison with data of net protons, we should consider more realistic quark number fluctuations and correlations in the studied rapidity window, which may be obtained by grand-canonical statistics of thermal quark system or by canonical statistics with Bernoulli trial selection of the specific window. We should also consider other effects related to the finite acceptance window such as the diffusion/blur of charges during the hadronic scatterings stage as well as that caused by resonance decay 
\cite{Fu:2009wy,Pan:2013xya,Westfall:2014fwa,Sakaida:2014pya,Nahrgang:2014fza,Kitazawa:2015ira,Braun-Munzinger:2016yjz,Ohnishi:2016bdf,Xu:2016qjd,Xu:2016qzd,Steinheimer:2016cir,Li:2017fhe,Li:2017via,He:2017zpg,Ling:2015yau}.
We will study these effects in this framework in the future.

\section{Summary}

\label{sec:summary}We developed a new statistical method to solve
the probability distribution for the numbers of baryons, antibaryons
and mesons formed in hadronization of a constituent quark and antiquark
system governed by the quark combination rule (QCR) in the quark combination
model developed by Shandong Group. We use a set of numbers $(N_{M},N_{B},N_{\bar{B}},N_{r},N_{\bar{r}})$
to classify the outcome of implementing the QCR for a queue of $N_{q}$
quarks and $N_{\bar{q}}$ antiquarks, where there are $N_{M}$ mesons,
$N_{B}$ baryons and $N_{\bar{B}}$ antibaryons formed by combination
but with $N_{r}$ quarks and $N_{\bar{r}}$ antiquarks left without
forming hadrons. The number of different ways of a given configuration
$(N_{M},N_{B},N_{\bar{B}},N_{r},N_{\bar{r}})$ is denoted as $F(N_{M},N_{B},N_{\bar{B}},N_{r},N_{\bar{r}})$.
We build the recursive relation for $F(N_{M},N_{B},N_{\bar{B}},0,0)$.
We adopt the generating function method to solve the recursive relation
and give the analytical expression of $F(N_{M},N_{B},N_{\bar{B}},0,0)$.
This method is much simpler than the previous one and is easy to generalize.
To accommodate the baryon yield in experimental data in relativistic
heavy-ion collisions, we consider a generalized combination rule (gQCR).
We give the solution of baryon and meson probability distribution
function under the new rule. Because the (anti)baryon production is
more suppressed in the gQCR, we find that the cumulant ratios of the
antibaryon number reach the Poisson statistics more rapidly in the
gQCR than in the QCR with increasing baryon number density. 

We studied the multiplicity fluctuation and correlation functions
for identified baryons directly produced in collisions. We also studied
correlation functions of final state hadrons including contributions
from resonance decays. As an illustrative example we consider a quark-antiquark
system having the property $S\sigma=z$ and $\kappa\sigma^{2}=1$
for the baryon quantum number, which is similar to the base line in
the grand-canonical ensemble in the statistical model. We calculate
the cumulant ratios in the quark combination model and find that $S\sigma$
for net protons in the model increases with decreasing collisional
energies, which is consistent with the experimental observation. More
interesting is that $\kappa\sigma^{2}$ for net protons has a nontrivial
energy behavior consistent with the data. %We argue that the cumulant ratios for net protons may not reflect the properties of net baryons which is directly related to the phase transition dynamics. 

%%%%%%%%%%%%%%%%%%%%%%%%%%%%%%%%%%%%%%%%%%%%%%%%%%%%%%%

\end{multicols}
\begin{appendix}

\section{Derivation of two properties for gQCR}

\label{sec:properties}We present two properties for $F(N_{M},N_{B},N_{\bar{B}},N_{r},N_{\bar{r}})$
in this appendix. 

Property 1. For $(N_{r},N_{\bar{r}})=(1,0),(0,1)$, $F(N_{M},N_{B},N_{\bar{B}},N_{r},N_{\bar{r}})$
can be expressed in terms of $F(N_{M}^{\prime},N_{B}^{\prime},N_{\bar{B}}^{\prime},0,0)$
with $N_{M}^{\prime}\leqslant N_{M}$, $N_{B}^{\prime}\leqslant N_{B}$
and $N_{\bar{B}}^{\prime}\leqslant N_{\bar{B}}$, 
\begin{align}
F(N_{M},N_{B},N_{\bar{B}},1,0) & =  \sum_{n_{M}=0}^{N_{M}}\sum_{n_{B}=0}^{N_{B}}\sum_{n_{\bar{B}}=0}^{N_{\bar{B}}}C_{n_{B},n_{\bar{B}}}^{10}(n_{M})2^{N_{M}-1-n_{M}+\delta_{n_{M},N_{M}}}F(n_{M},N_{B}-n_{B},N_{\bar{B}}-n_{\bar{B}},0,0),\nonumber \\
F(N_{M},N_{B},N_{\bar{B}},0,1) & =  \sum_{n_{M}=0}^{N_{M}}\sum_{n_{B}=0}^{N_{B}}\sum_{n_{\bar{B}}=0}^{N_{\bar{B}}}C_{n_{B},n_{\bar{B}}}^{01}(n_{M})2^{N_{M}-1-n_{M}+\delta_{n_{M},N_{M}}}F(n_{M},N_{B}-n_{B},N_{\bar{B}}-n_{\bar{B}},0,0),\label{eq:series-f10-01}
\end{align}
where the coefficients $C_{n_{B},n_{\bar{B}}}^{10}(n_{M})$ are given
by $C_{0,0}^{10}(n_{M})=1$ and 
\begin{align}
 & C_{a,b}^{10}(n_{M})=(-1)^{a+b}\sum_{j_{1}=n_{M}}^{N_{M}}\cdots\sum_{j_{a+b}=j_{a+b-1}}^{N_{M}}1,\quad\mathrm{for}\:b=a,a+1,\nonumber \\
 & C_{a,b}^{10}(n_{M})=0,\quad\mathrm{for}\:b\neq a,a+1,\label{eq:c10}
\end{align}
for $a+b>0$. The coefficients $C_{n_{B},n_{\bar{B}}}^{01}(n_{M})$
are given by $C_{0,0}^{01}(n_{M})=1$ and 
\begin{align}
 & C_{a,b}^{01}(n_{M})=(-1)^{a+b}\sum_{j_{1}=n_{M}}^{N_{M}}\cdots\sum_{j_{a+b}=j_{a+b-1}}^{N_{M}}1,\quad\mathrm{for}\:b=a,a-1,\nonumber \\
 & C_{a,b}^{01}(n_{M})=0,\quad\mathrm{for}\:b\neq a,a-1,\label{eq:c01}
\end{align}
for $a+b>0$. Note that the sums over $n_{B}$, $n_{\bar{B}}$ and
$n_{M}$ in Eq. (\ref{eq:series-f10-01}) continues until any of the
baryon, antibaryon and meson number become negative since $F(N_{M},N_{B},N_{\bar{B}},N_{r},N_{\bar{r}})=0$
for the case that any of $N_{M}$, $N_{B}$, $N_{\bar{B}}$, $N_{r}$
or $N_{\bar{r}}$ are negative. 

Property 2. For $(N_{r},N_{\bar{r}})=(2,0),(0,2)$, $F(N_{M},N_{B},N_{\bar{B}},N_{r},N_{\bar{r}})$
can be expressed in terms of $F(N_{M}^{\prime},N_{B}^{\prime},N_{\bar{B}}^{\prime},0,0)$
with $N_{M}^{\prime}\leqslant N_{M}$, $N_{B}^{\prime}\leqslant N_{B}$
and $N_{\bar{B}}^{\prime}\leqslant N_{\bar{B}}$, 
\begin{align}
F(N_{M},N_{B},N_{\bar{B}},2,0) & =  \sum_{n_{M}=0}^{N_{M}}\sum_{n_{B}=0}^{N_{B}}\sum_{n_{\bar{B}}=0}^{N_{\bar{B}}}C_{n_{B},n_{\bar{B}}}^{20}(n_{M})2^{N_{M}-n_{M}}F(n_{M},N_{B}-n_{B},N_{\bar{B}}-n_{\bar{B}},0,0),\nonumber \\
F(N_{M},N_{B},N_{\bar{B}},0,2) & =  \sum_{n_{M}=0}^{N_{M}}\sum_{n_{B}=0}^{N_{B}}\sum_{n_{\bar{B}}=0}^{N_{\bar{B}}}C_{n_{B},n_{\bar{B}}}^{02}(n_{M})2^{N_{M}-n_{M}}F(n_{M},N_{B}-n_{B},N_{\bar{B}}-n_{\bar{B}},0,0),\label{eq:series-f20-02}
\end{align}
where the coefficients are identical to Eqs. (\ref{eq:c10},\ref{eq:c01}):
$C_{n_{B},n_{\bar{B}}}^{20}(n_{M})=C_{n_{B},n_{\bar{B}}}^{10}(n_{M})$
and $C_{n_{B},n_{\bar{B}}}^{02}(n_{M})=C_{n_{B},n_{\bar{B}}}^{01}(n_{M})$. 

In order to prove the two properties in Eq. (\ref{eq:series-f10-01})
and Eq. (\ref{eq:series-f20-02}), we can solve $F(N_{M},N_{B},N_{\bar{B}},2,0)$
by substituting the first equation into the third one in Eq. (\ref{eq:gqcr-all}),
\begin{equation}
F(N_{M},N_{B},N_{\bar{B}},2,0)  =  \sum_{n_{M}=0}^{N_{M}}[F(n_{M},N_{B},N_{\bar{B}},0,0)-F(n_{M},N_{B},N_{\bar{B}}-1,0,2)]\times2^{N_{M}-n_{M}}.\label{eq:f20}
\end{equation}
In the same way, we can solve $F(N_{M},N_{B},N_{\bar{B}},0,2)$ by
substituting the second equation into the fourth one in Eq. (\ref{eq:gqcr-all})
\begin{equation}
F(N_{M},N_{B},N_{\bar{B}},0,2)  = \sum_{n_{M}=0}^{N_{M}}[F(n_{M},N_{B},N_{\bar{B}},0,0)-F(n_{M},N_{B}-1,N_{\bar{B}},2,0)]\times2^{N_{M}-n_{M}}.\label{eq:f02}
\end{equation}
From Eqs. (\ref{eq:f20},\ref{eq:f02}) we obtain Eq. (\ref{eq:series-f20-02}).
Then from Eq. (\ref{eq:series-f20-02}) and the last two equations
of (\ref{eq:gqcr-all}) we obtain Eq. (\ref{eq:series-f10-01}). 

%done

\section{Derivation of recursive equation Eq. (\ref{eq:recursive-f})}

\label{sec:recursive-f}We derive the recursive equation for $F(N_{M},N_{B},N_{\bar{B}},0,0)$
with the help of Eq. (\ref{eq:gqcr-all}). We start from Eq. (\ref{eq:f00-nm})
which also holds for gQCR. Using Eq. (\ref{eq:gqcr-all}), Eq. (\ref{eq:f00-nm})
can be rewritten as 
\begin{align}
F(N_{M},N_{B},N_{\bar{B}},0,0)  =G(N_{M},N_{B},N_{\bar{B}})-3G(N_{M}-1,N_{B},N_{\bar{B}})+G(N_{M}-2,N_{B},N_{\bar{B}}),\label{eq:f-g}
\end{align}
where $G$ is defined by 
\begin{equation}
G(N_{M},N_{B},N_{\bar{B}})  =  F(N_{M},N_{B},N_{\bar{B}},2,0)+F(N_{M},N_{B},N_{\bar{B}},0,2).\label{eq:gf}
\end{equation}
We need to define another auxiliary function 
\begin{align}
H(N_{M},N_{B},N_{\bar{B}}) & =  F(N_{M},N_{B}-1,N_{\bar{B}},2,0)+F(N_{M},N_{B},N_{\bar{B}}-1,0,2)\nonumber \\
 & =  2F(N_{M},N_{B},N_{\bar{B}},0,0)+2G(N_{M}-1,N_{B},N_{\bar{B}})-G(N_{M},N_{B},N_{\bar{B}})\nonumber \\
 & =  G(N_{M},N_{B},N_{\bar{B}})-4G(N_{M}-1,N_{B},N_{\bar{B}})+2G(N_{M}-2,N_{B},N_{\bar{B}}),\label{eq:hfg}
\end{align}
where we have used the first two equalities of Eq. (\ref{eq:gqcr-all})
to obtain second line and used Eq. (\ref{eq:f-g}) to obtain the last
line. On the other hand, we can also derive another form of $H(N_{M},N_{B},N_{\bar{B}})$
by applying the last two equalities and then the first two equalities
of Eq. (\ref{eq:hfg}) and finally applying Eq. (\ref{eq:gqcr-all}),
the result is 
\begin{align}
 H(N_{M},N_{B},N_{\bar{B}}) & =F(N_{M},N_{B}-1,N_{\bar{B}},1,0)+F(N_{M}-1,N_{B}-1,N_{\bar{B}},2,0) +F(N_{M},N_{B},N_{\bar{B}}-1,0,1) \nonumber \\
 & +F(N_{M}-1,N_{B},N_{\bar{B}}-1,0,2)\nonumber \\
 & =2H(N_{M}-1,N_{B},N_{\bar{B}})-G(N_{M},N_{B}-1,N_{\bar{B}}-1)+G(N_{M},N_{B}-1,N_{\bar{B}})\nonumber \\
 & -3G(N_{M}-1,N_{B}-1,N_{\bar{B}})+G(N_{M}-2,N_{B}-1,N_{\bar{B}})+G(N_{M},N_{B},N_{\bar{B}}-1)\nonumber \\
 & -3G(N_{M}-1,N_{B},N_{\bar{B}}-1)+G(N_{M}-2,N_{B},N_{\bar{B}}-1).\label{eq:h-2h-1}
\end{align}
From Eq. (\ref{eq:hfg}), we also obtain 
\begin{align}
 &   H(N_{M},N_{B},N_{\bar{B}})-2H(N_{M}-1,N_{B},N_{\bar{B}})\nonumber \\
 & =  G(N_{M},N_{B},N_{\bar{B}})-6G(N_{M}-1,N_{B},N_{\bar{B}}) +10G(N_{M}-2,N_{B},N_{\bar{B}})-4G(N_{M}-3,N_{B},N_{\bar{B}}).\label{eq:h-2h-2}
\end{align}
By equating Eq. (\ref{eq:h-2h-1}) and Eq. (\ref{eq:h-2h-2}) we derive
the recursive equation for $G$ and then for $F(N_{M},N_{B},N_{\bar{B}},0,0)$
through Eq. (\ref{eq:f-g}) as in Eq. (\ref{eq:recursive-f}). 

%done

\section{Derivation of generating functions for gQCR }

We multiply Eq. (\ref{eq:recursive-f}) by $x^{N_{M}}$ and sum over
$N_{M}\geqslant3$ (the equation is well defined for $N_{M}\geqslant3$),
we obtain 

\begin{align}
 & A(x;N_{B},N_{\bar{B}})-x^{2}F(2,N_{B},N_{\bar{B}},0,0)-xF(1,N_{B},N_{\bar{B}},0,0)-F(0,N_{B},N_{\bar{B}},0,0)\nonumber \\
 & =A(x;N_{B}-1,N_{\bar{B}})-x^{2}F(2,N_{B}-1,N_{\bar{B}},0,0)-xF(1,N_{B}-1,N_{\bar{B}},0,0)-F(0,N_{B}-1,N_{\bar{B}},0,0)\nonumber \\
 & +A(x;N_{B},N_{\bar{B}}-1)-x^{2}F(2,N_{B},N_{\bar{B}}-1,0,0)-xF(1,N_{B},N_{\bar{B}}-1,0,0)-F(0,N_{B},N_{\bar{B}}-1,0,0)\nonumber \\
 & -A(x;N_{B}-1,N_{\bar{B}}-1)+x^{2}F(2,N_{B}-1,N_{\bar{B}}-1,0,0)+xF(1,N_{B}-1,N_{\bar{B}}-1,0,0)\nonumber \\
 & +F(0,N_{B}-1,N_{\bar{B}}-1,0,0)+6x[A(x;N_{B},N_{\bar{B}})-xF(1,N_{B},N_{\bar{B}},0,0)-F(0,N_{B},N_{\bar{B}},0,0)]\nonumber \\
 & -3x[A(x;N_{B}-1,N_{\bar{B}})-xF(1,N_{B}-1,N_{\bar{B}},0,0)-F(0,N_{B}-1,N_{\bar{B}},0,0)]\nonumber \\
 & -3x[A(x;N_{B},N_{\bar{B}}-1)-xF(1,N_{B},N_{\bar{B}}-1,0,0)-F(0,N_{B},N_{\bar{B}}-1,0,0)]\nonumber \\
 & -10x^{2}[A(x;N_{B},N_{\bar{B}})-F(0,N_{B},N_{\bar{B}},0,0)]+x^{2}[A(x;N_{B}-1,N_{\bar{B}})\nonumber \\
 & -F(0,N_{B}-1,N_{\bar{B}},0,0)]+x^{2}[A(x;N_{B},N_{\bar{B}}-1)-F(0,N_{B},N_{\bar{B}}-1,0,0)]\nonumber \\
 & +4x^{3}A(x;N_{B},N_{\bar{B}}).\label{eq:rec-ax}
\end{align}
One can verify that a complete cancellation occurs for terms proportional
to $x^{2}$ and those proportional to $x$ after applying Eq. (\ref{eq:recursive-f})
for $N_{M}=2$ and $N_{M}=1$ respectively. The constant terms in $F$ reads
\begin{align}
I & =  F(0,N_{B},N_{\bar{B}},0,0)-F(0,N_{B}-1,N_{\bar{B}},0,0)-F(0,N_{B},N_{\bar{B}}-1,0,0)+F(0,N_{B}-1,N_{\bar{B}}-1,0,0)\nonumber \\
 & =  \delta_{N_{B}N_{\bar{B}},0}-\delta_{(N_{B}-1)N_{\bar{B}},0}-\delta_{N_{B}(N_{\bar{B}}-1),0}+\delta_{(N_{B}-1)(N_{\bar{B}}-1),0},
\end{align}
where we have used the initial value $F(0,N_{B},N_{\bar{B}},0,0)=\delta_{N_{B}N_{\bar{B}},0}$.
This means that if there are no mesons, baryons and antibaryons cannot
coexist, and if there are only quarks or antiquarks in the system
the number of different queues is 1. Then Eq. (\ref{eq:rec-ax}) is simplified as 
\begin{align}
A(x;N_{B},N_{\bar{B}}) & =\left(6x+4x^{3}-10x^{2}\right)A(x;N_{B},N_{\bar{B}}) +\left(1-3x+x^{2}\right)\left[A(x;N_{B}-1,N_{\bar{B}})+A(x;N_{B},N_{\bar{B}}-1)\right]\nonumber \\
 & -A(x;N_{B}-1,N_{\bar{B}}-1) +\delta_{N_{B}N_{\bar{B}},0}-\delta_{(N_{B}-1)N_{\bar{B}},0}-\delta_{N_{B}(N_{\bar{B}}-1),0}+\delta_{(N_{B}-1)(N_{\bar{B}}-1),0}.\label{eq:rec-ax1}
\end{align}
We now multiply Eq. (\ref{eq:rec-ax1}) by $y^{N_{B}}z^{N_{\bar{B}}}$ and take a sum over $N_{B}\geqslant1$ and $N_{\bar{B}}\geqslant1$ to obtain 
\begin{align}
 & \left(1-6x+10x^{2}-4x^{3}\right)A(x,y,z)\nonumber \\
 & =\left(1-3x+x^{2}\right)\sum_{N_{\bar{B}}=1}^{\infty}\sum_{N_{B}=1}^{\infty}y^{N_{B}}z^{N_{\bar{B}}}\nonumber \left[A(x;N_{B}-1,N_{\bar{B}})+A(x;N_{B},N_{\bar{B}}-1)\right]\nonumber  -\sum_{N_{\bar{B}}=1}^{\infty}\sum_{N_{B}=1}^{\infty}A(x;N_{B}-1,N_{\bar{B}}-1)y^{N_{B}}z^{N_{\bar{B}}}\nonumber \\
 & +\sum_{N_{\bar{B}}=1}^{\infty}\sum_{N_{B}=1}^{\infty}\left[\delta_{N_{B}N_{\bar{B}},0}-\delta_{(N_{B}-1)N_{\bar{B}},0}-\delta_{N_{B}(N_{\bar{B}}-1),0}\right. \left.+\delta_{(N_{B}-1)(N_{\bar{B}}-1),0}\right]y^{N_{B}}z^{N_{\bar{B}}},\label{eq:recursive-a}
\end{align}
To simplify the above equation, we use
\begin{align}
 & \sum_{N_{\bar{B}}=1}^{\infty}\sum_{N_{B}=1}^{\infty}A(x;N_{B}-a,N_{\bar{B}}-b)y^{N_{B}}z^{N_{\bar{B}}}\nonumber \\
 & =\delta_{a,1}\delta_{b,1}yzA(x,y,z) +\delta_{a,0}\delta_{b,1}z\left[A(x,y,z)-\sum_{N_{\bar{B}}=0}^{\infty}A(x;0,N_{\bar{B}})z^{N_{\bar{B}}}\right] +\delta_{a,1}\delta_{b,0}y\left[A(x,y,z)-\sum_{N_{B}=0}^{\infty}A(x;N_{B},0)y^{N_{B}}\right]\nonumber \\
 & +\delta_{a,0}\delta_{b,0}\left[A(x,y,z)-\sum_{N_{B}=0}^{\infty}A(x;N_{B},0)y^{N_{B}}\right. \left.-\sum_{N_{\bar{B}}=0}^{\infty}A(x;0,N_{\bar{B}})z^{N_{\bar{B}}}+A(x;0,0)\right],
\end{align}
and 
\begin{align}
-yz & =\sum_{N_{\bar{B}}=1}^{\infty}\sum_{N_{B}=1}^{\infty}\left[\delta_{N_{B}N_{\bar{B}},0}-\delta_{(N_{B}-1)N_{\bar{B}},0}\right. \left.-\delta_{N_{B}(N_{\bar{B}}-1),0}+\delta_{(N_{B}-1)(N_{\bar{B}}-1),0}\right]y^{N_{B}}z^{N_{\bar{B}}},\label{eq:recursive-a2}
\end{align}
as well as Eqs. (\ref{eq:axyp1}) and (\ref{eq:axyz}), we finally obtain Eq. (\ref{eq:gen-func-axyz}).

%done

\section{Derivation of Eq. (\ref{eq:re_gqcr}) in gQCR}

\label{sec:coefficient-gqcr-1}From Eq. (\ref{eq:gen-func-axyz})
we can extract $F(N_{M},N_{B},N_{\bar{B}},0,0)$, the coefficient
of $x^{N_{M}}y^{N_{B}}z^{N_{\bar{B}}}$ in the polynomial expansion
of $A(x,y,z)$. To this end we rewrite $A(x,y,z)$ as
\begin{align}
   & A(x,y,z)\nonumber \\
 & =  \frac{1-4x+4x^{2}-yz}{1-6x+10x^{2}-4x^{3}}\sum_{i=0}^{\infty}\left[\frac{1-3x+x^{2}}{1-6x+10x^{2}-4x^{3}}\left(y+z\right)-\frac{1}{1-6x+10x^{2}-4x^{3}}yz\right]^{i}\nonumber \\
 & =  \frac{1-4x+4x^{2}-yz}{1-6x+10x^{2}-4x^{3}}\sum_{i=0}^{\infty}\sum_{j+k+l=i}\left(\begin{array}{c}
i\\
j,\ k,\ l
\end{array}\right)y^{j+l}z^{k+l} \times\left(\frac{1-3x+x^{2}}{1-6x+10x^{2}-4x^{3}}\right)^{j+k}\left(\frac{-1}{1-6x+10x^{2}-4x^{3}}\right)^{l}\nonumber \\
 & =  \frac{1-4x+4x^{2}-yz}{1-6x+10x^{2}-4x^{3}}\sum_{i=0}^{\infty}\sum_{j=0}^{i}\sum_{k=0}^{i}\left(\begin{array}{c}
i\\
j,\ k,\ i-j-k
\end{array}\right)y^{i-k}z^{i-j}\times\left(\frac{1-3x+x^{2}}{1-6x+10x^{2}-4x^{3}}\right)^{j+k}\left(\frac{-1}{1-6x+10x^{2}-4x^{3}}\right)^{i-j-k},
\end{align}
where we have used the multinomial theorem 
\begin{equation}
{\displaystyle (w_{1}+w_{2}+\cdots+w_{m})^{n}}=\sum_{k_{1}+k_{2}+\cdots+k_{m}=n}{n \choose k_{1},\ k_{2},\ \ldots,\ k_{m}}\prod_{t=1}^{m}w_{t}^{k_{t}},
\end{equation}
with multinomial coefficients 
\begin{equation}
{n \choose k_{1},\ k_{2},\ \ldots,\ k_{m}}=\frac{n!}{k_{1}!k_{2}!\cdots k_{m}!}.
\end{equation}

Then we can extract the coefficient of $y^{N_{B}}z^{N_{\bar{B}}}$
as 
\begin{equation}
C(y^{N_{B}}z^{N_{\bar{B}}})=C_{1}+C_{2}+C_{3}+C_{4},\label{eq:coeff-yz-1}
\end{equation}
where $C_{1,2,3,4}$ are given by 
\begin{align}
C_{1} & =  \frac{1}{1-6x+10x^{2}-4x^{3}}\sum_{i=0}^{\infty}\left(\begin{array}{c}
i\\
i-N_{B},\ i-N_{\bar{B}},\ N_{B}+N_{\bar{B}}-i
\end{array}\right) \left(\frac{1-3x+x^{2}}{1-6x+10x^{2}-4x^{3}}\right)^{2i-N_{B}-N_{\bar{B}}}\left(\frac{-1}{1-6x+10x^{2}-4x^{3}}\right)^{N_{B}+N_{\bar{B}}-i}\nonumber \\
C_{2} & =  \frac{-4x}{1-6x+10x^{2}-4x^{3}}\sum_{i=0}^{\infty}\left(\begin{array}{c}
i\\
i-N_{B},\ i-N_{\bar{B}},\ N_{B}+N_{\bar{B}}-i
\end{array}\right) \left(\frac{1-3x+x^{2}}{1-6x+10x^{2}-4x^{3}}\right)^{2i-N_{B}-N_{\bar{B}}}\left(\frac{-1}{1-6x+10x^{2}-4x^{3}}\right)^{N_{B}+N_{\bar{B}}-i}\nonumber \\
C_{3} & =  \frac{4x^{2}}{1-6x+10x^{2}-4x^{3}}\sum_{i=0}^{\infty}\left(\begin{array}{c}
i\\
i-N_{B},\ i-N_{\bar{B}},\ N_{B}+N_{\bar{B}}-i
\end{array}\right)  \left(\frac{1-3x+x^{2}}{1-6x+10x^{2}-4x^{3}}\right)^{2i-N_{B}-N_{\bar{B}}}\left(\frac{-1}{1-6x+10x^{2}-4x^{3}}\right)^{N_{B}+N_{\bar{B}}-i}\nonumber \\
C_{4} & =  -\frac{1}{1-6x+10x^{2}-4x^{3}}\sum_{i=0}^{\infty}\left(\begin{array}{c}
i\\
i-N_{B}+1,\ i-N_{\bar{B}}+1,\ N_{B}+N_{\bar{B}}-2-i
\end{array}\right)\nonumber \\
 &   \times\left(\frac{1-3x+x^{2}}{1-6x+10x^{2}-4x^{3}}\right)^{2i-N_{B}-N_{\bar{B}}+2}\left(\frac{-1}{1-6x+10x^{2}-4x^{3}}\right)^{N_{B}+N_{\bar{B}}-2-i}.
\end{align}
Note that we have expressed the summation indices $j$ and $k$ in
terms of $N_{B}$ and $N_{\bar{B}}$ for a given summation index $i$.
The sum over $i$ involves a finite number of terms. For example,
in $C_{1}$, we have $N_{B}+N_{\bar{B}}\geqslant i\geqslant\mathrm{Max}(N_{B},N_{\bar{B}})$,
since any terms in the summation with the factorial of a negative
integer in the denominator are vanishing. 

We can factorize the following two polynomials as 
\begin{align}
 &1-3x+x^{2}  =  \left(1-\frac{3+\sqrt{5}}{2}x\right)\left(1-\frac{3-\sqrt{5}}{2}x\right),\\
    & 6x+10x^{2}-4x^{3}  =  6x\left(1-x\right)\left(1-\frac{2}{3}x\right),
\end{align}
and expand $C_{1}$, $C_{2}$, $C_{3}$, $C_{4}$ into a power series
of $x$ with the help of the binomial theorem. Then we can extract
the coefficient of $x^{N_{M}}$ in $C(y^{N_{B}}z^{N_{\bar{B}}})$.
This gives the coefficient of $x^{N_{M}}y^{N_{B}}z^{N_{\bar{B}}}$
in $A(x,y,z)$ or $F(N_{M},N_{B},N_{\bar{B}},0,0)$ as the sum of
the following four terms 
\begin{align}
I_{1} & =  \sum_{i=0}^{\infty}\left(-1\right)^{N_{B}+N_{\bar{B}}-i}\left(\begin{array}{c}
i\\
i-N_{B},\ i-N_{\bar{B}},\ N_{B}+N_{\bar{B}}-i
\end{array}\right)  \sum_{j+k+l+m+n=N_{M}}\left(\begin{array}{c}
2i-N_{B}-N_{\bar{B}}\\
j
\end{array}\right)\left(\begin{array}{c}
2i-N_{B}-N_{\bar{B}}\\
k
\end{array}\right)\nonumber \\
 &   \times\left(-\frac{3+\sqrt{5}}{2}\right)^{j}\left(-\frac{3-\sqrt{5}}{2}\right)^{k}\left(\begin{array}{c}
i+l\\
l
\end{array}\right)6^{l}\left(\begin{array}{c}
l\\
m
\end{array}\right)\left(-1\right)^{m}\left(\begin{array}{c}
l\\
n
\end{array}\right)\left(-\frac{2}{3}\right)^{n}\nonumber \\
I_{2} & =  -4\sum_{i=0}^{\infty}\left(-1\right)^{N_{B}+N_{\bar{B}}-i}\left(\begin{array}{c}
i\\
i-N_{B},\ i-N_{\bar{B}},\ N_{B}+N_{\bar{B}}-i
\end{array}\right) \sum_{j+k+l+m+n=N_{M}-1}\left(\begin{array}{c}
2i-N_{B}-N_{\bar{B}}\\
j
\end{array}\right)\left(\begin{array}{c}
2i-N_{B}-N_{\bar{B}}\\
k
\end{array}\right)\nonumber \\
 &   \times\left(-\frac{3+\sqrt{5}}{2}\right)^{j}\left(-\frac{3-\sqrt{5}}{2}\right)^{k}\left(\begin{array}{c}
i+l\\
l
\end{array}\right)6^{l}\left(\begin{array}{c}
l\\
m
\end{array}\right)\left(-1\right)^{m}\left(\begin{array}{c}
l\\
n
\end{array}\right)\left(-\frac{2}{3}\right)^{n}\nonumber \\
I_{3} & =  4\sum_{i=0}^{\infty}\left(-1\right)^{N_{B}+N_{\bar{B}}-i}\left(\begin{array}{c}
i\\
i-N_{B},\ i-N_{\bar{B}},\ N_{B}+N_{\bar{B}}-i
\end{array}\right)  \sum_{j+k+l+m+n=N_{M}-2}\left(\begin{array}{c}
2i-N_{B}-N_{\bar{B}}\\
j
\end{array}\right)\left(\begin{array}{c}
2i-N_{B}-N_{\bar{B}}\\
k
\end{array}\right)\nonumber \\
 &   \times\left(-\frac{3+\sqrt{5}}{2}\right)^{j}\left(-\frac{3-\sqrt{5}}{2}\right)^{k}\left(\begin{array}{c}
i+l\\
l
\end{array}\right)6^{l}\left(\begin{array}{c}
l\\
m
\end{array}\right)\left(-1\right)^{m}\left(\begin{array}{c}
l\\
n
\end{array}\right)\left(-\frac{2}{3}\right)^{n}\nonumber \\
I_{4} & =  -\sum_{i=0}^{\infty}\left(-1\right)^{N_{B}+N_{\bar{B}}-i}\left(\begin{array}{c}
i\\
i-N_{B}+1,\ i-N_{\bar{B}}+1,\ N_{B}+N_{\bar{B}}-2-i
\end{array}\right)  \sum_{j+k+l+m+n=N_{M}}\left(\begin{array}{c}
2i-N_{B}-N_{\bar{B}}+2\\
j
\end{array}\right)\left(\begin{array}{c}
2i-N_{B}-N_{\bar{B}}+2\\
k
\end{array}\right)\nonumber \\
 &   \times\left(-\frac{3+\sqrt{5}}{2}\right)^{j}\left(-\frac{3-\sqrt{5}}{2}\right)^{k}\left(\begin{array}{c}
i+l\\
l
\end{array}\right)6^{l}\left(\begin{array}{c}
l\\
m
\end{array}\right)\left(-1\right)^{m}\left(\begin{array}{c}
l\\
n
\end{array}\right)\left(-\frac{2}{3}\right)^{n},\label{eq:re_gqcr_all}
\end{align}
which give the final result of gQCR in Sec. \ref{sec:new_QCR}.

\end{appendix}

\clearpage
\end{CJK*}
\end{document}